\documentstyle{article}

\def\bichpr{\hoffset=-20truemm
\voffset=-30truemm
\textwidth=16 truecm
\textheight=24 truecm }

\bichpr
\begin{document}
\title{Two and three-fermion 3D equations deduced from Bethe-Salpeter
equations.} 
\author{ J. Bijtebier\thanks{Senior Research Associate at the
 Fund for Scientific Research (Belgium).}\\
 Theoretische Natuurkunde, Vrije Universiteit Brussel,\\
 Pleinlaan 2, B1050 Brussel, Belgium.\\ Email: jbijtebi@vub.ac.be}
\maketitle
\begin{abstract}   \noindent
  We write a 3D equation for three fermions by combining the three
two-body potentials obtained in 3D reductions (based on a series expansion
around a relative-energy fixing "approximation" of the free
propagators) of the corresponding two-fermion Bethe-Salpeter equations to
equivalent 3D equations, putting the third fermion on its
positive-energy mass shell. In this way, the cluster-separated limits are
exact, and the Lorentz invariance / cluster separability requirement is
automatically satisfied, provided no supplementary approximation, like the
Born approximation, is made. The use of positive free-energy projectors in
the chosen reductions of the two-fermion Bethe-Salpeter  equations prevents
continuum dissolution in our 3D three-fermion equation. The potentials are
hermitian below the inelastic threshold and depend only slowly on the total
three-fermion energy.  The one high-mass limits are approximately exact. \par
This "hand-made" three-fermion 3D equation is also obtained by starting with
an approximation of the three-fermion Bethe-Salpeter
equation, in which the three-body kernel is neglected and the two-body
kernels approached by positive-energy instantaneous expressions, with the
spectator fermion on the mass shell. The neglected terms are then
transformed into corrections to the 3D equation, in three steps implying
each a series expansion. The result is of course complicated, but the
lowest-order contributions of these correction terms to the energy spectrum
remain manageable.\par
 We also present some other 3D reduction procedures and
compare them to our's: use of Sazdjian's covariant approximation of the
free propagator, 3D reductions performed by a series expansion around 
instantaneous approximations of the kernels instead of "approximations" of
the propagators,  Gross' spectator model.
\end{abstract}
electrodynamics.\par  Relativistic bound states. Relativistic
wave equations. \\\\
 \newpage
\tableofcontents

\section{Introduction} In the treatment of the three-body problem, there
is a large gap between nonrelativistic quantum mechanics (Schr\"odinger
equation) and relativistic quantum field theory (Bethe-Salpeter equation
\cite{1,2}). Starting from the Schr\"odinger equation one can of course
replace the free part of the hamiltonian by its relativistic form.
Starting from the Bethe-Salpeter equation, one can try to eliminate the
two relative time variables to get finally a Schr\"odinger equation with
a compact potential plus a lot of correction terms of various origins.
Instead of the Schr\"odinger equation, one can use the Faddeev equations.
These equations can be derived from the Schr\"odinger equation, or, in a
more general form, from the Bethe-Salpeter equation. Faddeev equations
for the transition operator give the various scattering and reaction
matrix elements and the poles (to be computed by basically
nonperturbative methods) of this operator in the total energy give the
spectrum of the three-body bound states.\par Schr\"odinger's equation is
"relatively easy" to solve, but this zero-order approximation does not
reflect several important properties and symmetries of the studied
physical system. These could be recovered, in principle, by incorporating
higher-order contributions (in general an infinity of them).\par As an
intermediate step between nonrelativistic quantum mechanics and quantum
field theory we shall search for a 3D equation built with the sum of the
relativistic free hamiltonians plus two-body and perhaps also three-body
interaction potentials. Such equations are closely related to the systems
of coupled Dirac or Klein-Gordon equations of constraint theory [3-7],
which are still 4D equations but exhibit a simplified dependence in the
relative times, which could be completely eliminated to get a single 3D
equation. Our 3D equation will be a strategic link in a chain of
approximations between a manageable equation (such as Schr\"odinger's
equation) and a full 3D equivalent (at least for what concerns the bound
states spectrum) of the original Bethe-Salpeter equation. We shall try to
keep a maximum of properties and symmetries of the original
Bethe-Salpeter equation in our 3D equation. For the properties which we
shall be unable to satisfy exactly in our 3D equation, we expect that the
inclusion of
higher-order correction terms will progressively improve the situation.
We shall thus try to make our 3D equation satisfy the following list of
requirements:\par\hfill\break\indent --- Correct nonrelativistic
limit.\par --- Lorentz invariance and cluster separability. It is always
possible to render an equation Lorentz invariant by working in the
general rest frame (center of mass reference frame) and by building
invariants with the total 4-momentum vector, although the result may be
unelegant and artificial. The Lorentz invariance requirement becomes a
tool when combined with the cluster separability requirement: when all
mutual interactions are "switched off", we must get a set of three free
Dirac equations. This total separability can  easily  be obtained by using
as hamiltonian the sum of three free Dirac hamiltonians and interaction
terms. The real difficulty appears when only the interactions with
fermion 3 (for example) are switched off. If we want a full cluster
separability, the resulting equation for the (12) cluster can not refer
to the global center of mass frame anymore, as the momentum of fermion 3
enters in the definition of this frame. \par Lorentz invariance and
cluster separability can be explicit or implicit (via rearrangements).
The best known example of implicit Lorentz invariance is a free Dirac
equation solved with respect to the energy: it becomes explicitly
covariant by multiplication with the $\,\beta\,$ matrix. Other implicit
Lorentz invariances are not that trivial. For example, the 3D reductions
of a Bethe-Salpeter equation are implicitly covariant, provided the
series generated by this reduction is not truncated.\par In fact, the
homogeneous Bethe-Salpeter equation from which we shall start does not
obey cluster separability: this equation is valid for three-fermion bound
states only, i.e. for a total energy below all the continua. Our 3D
equation will thus be used in the computation of three-fermion bound
states. A priori, it will not necessarily give also the correct
scattering amplitudes or obey cluster separability, even with all
higher-order terms included. We shall nevertheless try to get a 3D
equation which is also valid in the continuum region.\par 
 --- Hermiticity and total energy independence of the interaction terms.
In the two-body problem, nonhermitian interaction terms can be hermitian
with respect to a modified scalar product, or made hermitian via a
rearrangement of the equation. In the three-body problem these
rearrangements could be more complicated. The hermiticity and the
independence on the total energy are linked features, as one of these is
often achieved at the expense of the other one. Energy depending
interaction terms destroy some of the advantages of the use of an
hermitian hamiltonian, such as the mutual orthogonality of the solutions,
and leads to modify the usual perturbation calculation methods. The 3D
potentials deduced from field theory are generally energy dependent, at
least in the higher-order terms. We shall require  hermiticity and
energy-independence (or slow energy  dependence) below the inelastic
threshold in the lowest order terms at least.\par  --- Correct heavy mass
limits. When the mass of one of the fermions becomes infinite, its
presence must be translated in the equations by a potential (Coulombian
in QED) acting on the other fermions. For the two-body problem the
"one-body limit" in QED is indeed a Dirac-Coulomb equation or a
rearrangement of it (for example when a projection operator is introduced
in order to avoid continuum dissolution - see below). In the
Dirac-Coulomb equation, the Coulomb potential is already given by the
limit of the Born term, as the higher-order crossed and ladder terms
cancel mutually at the one-body limit. In contrast, the one-body limits
of the rearrangements contain contributions from all terms [8-12].\par
--- Perturbative approach. It must be possible to start with a manageable
approximate equation and indefinitely improve the approximation of the
measurable quantities (with respect to the uncalculated predictions of
the here assumed exact Bethe-Salpeter equation) by adding higher-order
contributions.\par --- Solution of the continuum dissolution problem. In
the relativistic equations for several relativistic particles, the
physical bound states are degenerate with a continuum of states combining
asymptotically free particles with opposite energy signs. This often
neglected fact forbids the building of normalizable solutions in the
$\,N\!>\!2$-body problem (including the two-body plus potential problem.
In the pure two-body problem the mixing is prevented by the conservation
of the total momentum). The usual solution consists in including
positive-energy projection operators into the zero-order propagator
[13-16]. The modified equations must of course continue to satisfy the
other requirements, like the Lorentz invariance / cluster separability
requirement.\par\hfill\break\indent

Among these requirements, the absence of continuum dissolution is a technical
necessity, if we want to get normalizable bound states. The correct
nonrelativistic limit is of course a must. The possibility of improving the
chosen approximation could enable us to estimate the precision of this
approximation and possibly to satisfy the other requirements up to the
considered order at least.

\par\hfill\break\indent
 Section 2 is devoted to the two-fermion problem, revisited in order to
define the notations and present the building blocks and
cluster-separated limits of our future three-fermion equations.  In
section 3 we write directly a cluster separable 3D equation, by combining
the relativistic free hamiltonians and the 3D potentials obtained in the
reduction of the two-fermion Bethe-Salpeter equations, putting in each 3D
potential the third fermion on the mass shell. Switching off two of
the three mutual interaction potentials gives then a free Dirac equation
for a spectator fermion plus the two-body equation which would be obtained
directly by 3D-reducing the two-fermion Bethe-Salpeter equation, for the two
interacting fermions. This  insures the Lorentz invariance /
cluster separability property. Positive-energy projectors solve the continuum
dissolution problem. Our 3D equation can also be obtained from the
three-fermion Bethe-Salpeter equation, by neglecting the three-body kernel
and replacing the two-body kernels by positive-energy instantaneous (i.e.
independent of the relative energies) approximations, equivalent at the
cluster-separated limits. The remaining dependence (in the free
propagators) of the approximated Bethe-Salpeter equation in the relative
energies can then be eliminated by performing closed path integrals. This
derivation of our 3D equation from an approximated Bethe-Salpeter equation
enabled us to transform the neglected
terms, known at the  Bethe-Salpeter equation level, into correction terms to
the 3D equation. We achieved this by performing three consecutive expansions of
the correction terms: the three-fermion  Bethe-Salpeter
equation is first transformed into a set of three coupled equations for three
 wave
functions depending each on one two-fermion total energy, then into a set of
three 3D equations and finally into a single 3D equation. The resulting
expression for the 3D equation is rather complicated, as it combines three
series expansions, but we use it to write a manageable expression of the
first-order corrections to the energy spectrum.  The contributions of the
residues of the poles
of the propagators to this energy shift are easy to compute in terms of the
Bethe-Salpeter kernels with two of the three fermions on their mass shell, in
the spirit of Gross' spectator model
\cite{12,13}. Besides, we have also the contributions of the singularities
of the kernels themselves to the closed path integrals. These contributions
can not be a priori neglected.
In section 4, we examine some other possible choices of the "approximation"
of the free propagator, including Sazdjian's covariant propagator
\cite{18,7,14}. In combination with a covariant substitute of the projector on
the positive free-energy states, this propagator leads to a 3D potential
given by a series in which each term is separately covariant, so that we can
truncate the series without breaking the covariance. In section 5, we examine
an alternative 3D reduction method built around instantaneous approximations
of the Bethe-Salpeter kernels. The non-hermitian resulting 3D potential can be
made hermitian either by performing a second series expansion, easily
combinable with the series expansion generated by the 3D reduction, or by
adaptating the starting approximation of the Bethe-Salpeter kernel (Phillips
and Wallace's method \cite{21}). It is interesting to note that Phillips
and Wallace's method and the propagator-based 3D reduction method give both
the correct scattering operator (a bonus not implied by the starting
homogeneous Bethe-Salpeter equation, designed for bound states), while the
ordinary kernel-based 3D reduction does not. Section 6 is devoted to a
comparison of our "hand-made" 3D equation with Gross'  spectator model's
equations
\cite{12,13} in Faddeev's formalism. In this model, the dependence of the
two-fermion transition operators in the relative energies is taken into account
by putting different fermions on their positive-energy mass shells, according
to the neighbours of each two-fermion transition operator in the expansion
giving the three-fermion transition operator. Each of our two-fermion
transition operators is everywhere approximated in the same way, so that our
Faddeev equations are much simpler and can be transformated into our wave
equation. The difference between both models corresponds to
higher-order contributions, which we recover in our calculation of the
higher-order terms, together with the supplementary contributions, not to be a
priori neglected, of the singularities of the kernels to the closed path
integrals in the relative energies complex planes. In section 7, we compare the
one-high mass limits ("two-body limits") of our three-fermion equation with our
previous two-fermion in an external potential equation \cite{11}.  Section 8
is devoted to conclusions.
   
\section{The two-fermion problem.}
\subsection{Notations.}
\noindent We shall write the Bethe-Salpeter equation for the bound states
 of two fermions \cite{1} as 
\begin{equation}\Phi = G^0 K \Phi,    \label{1}\end{equation}   where
$\Phi$ is the Bethe-Salpeter  amplitude, function of the positions
$x_1,x_2$ or of the momenta 
$p_1,p_2$ of the fermions, according to the representation chosen. The
operator $K$  is the Bethe-Salpeter kernel, given as a factor of the kernel
of an integral equation in momentum space by the sum of the irreducible
two-fermion Feynman graphs. The operator
$G^0$ is the free propagator, given by the product
$G^0_1G^0_2$ of the two individual fermion propagators
\begin{equation}G^0=G^0_1G^0_2,\qquad G^0_i = {1 \over p_{i0}-h_i+i\epsilon
h_i}\,\beta_i = {p_{i0}+h_i\over p_i^2-m_i^2+i\epsilon}\,\beta_i 
\label{2}\end{equation}  where the $h_i$ are the Dirac free hamiltonians
\begin{equation}h_i = \vec \alpha_i\, . \vec p_i + \beta_i\, m_i\qquad
(i=1,2).  \label{3}\end{equation}  The Bethe-Salpeter kernel $\,K \,$
should contain charge renormalization and vacuum polarization graphs,
while the propagators $\, G^0_i\,$ should contain self-energy terms
(which can be transferred to $\, K\,$ \cite{14,15} ]). In this work, we
consider only the free fermion propagators in   $\,G^0_i\,$ and the
"skeleton" graphs in $\,K$, and we hope that the inclusion of the various
corrections would not change our conclusions.\par\noindent  We shall
define the total (or external, CM, global) and relative (or internal)
variables:
\begin{equation}X = {1 \over 2} (x_1 + x_2)\ , \qquad P = p_1 + p_2\ , 
\label{4}\end{equation} 
\begin{equation}x = x_1 - x_2\ , \qquad p = {1 \over 2} (p_1 - p_2).
\label{5}\end{equation}  and give a name to the corresponding
combinations of the free hamiltonians:
\begin{equation}S = h_1 + h_2\ , \quad s = {1 \over 2} (h_1 -
h_2).\label{6}\end{equation}  We know that, at the no-interaction limit,
the system is described by a pair of free Dirac equations:
\begin{equation} (p_{10}-h_1)\Psi=0, \qquad (p_{20}-h_2)\Psi=0, 
\label{7}\end{equation}  where $\,\Psi \,$ depends on 
$\,x_1,x_2.\,$ Let us also write their iterated version
\begin{equation} (p_{10}^2-E_1^2)\Psi=0, \qquad (p_{20}^2-E_2^2)\Psi=0 
\label{8}\end{equation}  with
\begin{equation} E_i=\sqrt{h_i^2}=(\vec p_i^2+m_i^2)^{1\over 2}.  
\label{9}\end{equation}  Interesting combinations can be obtained from
the sum and differences of the equations (\ref{7}) or of the iterated
equations (\ref{8}):
\begin{equation}(P_0-S)\Psi=0, \qquad (p_0-s)\Psi=0, 
\label{10}\end{equation} 
\begin{equation}H^0\Psi=0, \qquad (p_0-\mu)\Psi=0
\label{11}\end{equation}  with
\begin{equation}H^0=2[(p_1^2-m_1^2)+(p_2^2-m_2^2)]_{p_0=\mu}\,\,=\,
\,P_0^2-2(E_1^2+E_2^2)+4\mu^2, \label{12}\end{equation} 
\begin{equation}\mu ={1\over 2P_0}(E_1^2-E_2^2)={1\over
2P_0}(h_1^2-h_2^2)={sS\over P_0}.  \label{13}\end{equation}
The choice (\ref{4},\ref{5}) of the variables $\,X\,$ and $\,p\,$ was
partially arbitrary: other linear combinations are possible, provided the
canonical commutation relations are satisfied. In the nonrelativistic
framework, the choice of $\,\vec X\,$ is weighted by the masses, and
$\,\vec p\,$ is chosen accordingly. A frequent choice in the relativistic
framework is a weighting by the free energies, which leads to a simpler
complementary equation $\,p_0\Psi\!=\!0\,$ instead of (\ref{11}), but has
the drawback that the coefficients of the weighting are not constant. We
think that our choice (\ref{4},\ref{5}) leads to simpler expressions,
even if taking the nonrelativistic limit is less straightforward. \par
Our free equations (\ref{10}) are eigenvalue equations, the energies $\,P_0\,$
and
$\,p_0\,$ being the eigenvalues and $\,S\,$ and $\,s\,$ the corresponding
operators. Equation (\ref{11}) with $\,H^0\,$ is a fourth-order equation in
$\,P_0,\,$ with four solution for a given set of values of the spatial
momenta. Equation (\ref{11}) with $\,p_0\,$ is the eigenvalue equation
of an operator $\,\mu,\,$ depending itself on $\,P_0.\,$ In the
Bethe-Salpeter equation (\ref{1}), $\,P_0\,$ remains a constant, but
$\,p_0\,$ becomes an operator in position space, or a variable in
momentum space. The 3D reduction will consist in eliminating this
variable by performing an expansion around a fixed combination of the other ones
($\,s\,$ or
$\,\mu\,$ in the following).

\subsection{3D reduction of the two-fermion Bethe-Salpeter equation.} In
the zero-order approximation, the free propagator $G^0$ will be replaced
by a carefully chosen expression
$G^\delta$, combining a constraint like $\delta(p_0\! -\!\mu)$ fixing the
relative energy, and a global 3D propagator like $-2i\pi (P_0\! -\!
S)^{-1}\beta_1\beta_2$. The argument of the $\delta$ and the inverse of
the propagator should be combinations of the operators used in the free
equations (at last approximately and for the positive-energy solutions).
There exists an infinity of possible combinations [8, 12, 20, 22-34]. The
best choice depends on the quantities one wants to compute (energy of the
lowest state, hyperfine splitting, recoil of a nucleus, etc...) and on
the properties one wants to preserve exactly in the first approximation
(cluster separability, Lorentz invariance, heavy mass limits,
 charge conjugation symmetry...). All choices would be equivalent if all
correction terms could be computed, but this is of course impossible.
\par We shall see below that the reduced wave function $\Psi$ (from which
the relative time-energy degree of freedom can be trivially eliminated)
is given by
\begin{equation}\Psi=G^\delta (G^0)^{-1}\Phi. \label{14}\end{equation} 
The choice of the
constraint fixes the 3D hypersurface ($p_0\! =\!\mu$ for example)
on which we want to work. The remaining of
$G^\delta$ is a purely 3D operator, the different choices of which result
in different 3D operators applied on a common basic 3D wave function
 and in different rearrangements of a common reduction series
giving the 3D potential. The various propagator-based 3D reductions of
the literature can thus be classified according to the constraint they
use. Once this constraint chosen, we can only write different equivalent
forms (or sometimes projections) of the same 3D equation. It is the
unavoidable truncation of the reduction series which makes the difference
(numeric, if we simply want to compute predictions of field theory; more
fundamental, if we want to write constraint theory equations with QCD
inspired potentials).\par Two natural choices for the constraint are
$\delta(p_0\! -\! s)$, based on the first-order equations (\ref{10}) and 
$\delta(p_0\! -\!\mu)$, based on the second-order equations (\ref{11}).
In $\delta(p_0\! -\! s)$, $\,s\,$ is an operator which can be
diagonalized in momentum space using the four projectors on the subspaces
corresponding to the different signs of $\,h_1,h_2\,$ (see below). By
contrast, $\,\mu\,$ is a momentum-depending number, which reduces to
$\,(m_1^2\!-\!m_2^2)/2P_0\,$ in the two-fermion center of mass frame
$\,\vec P\!=\!0\,$ which can be defined in the pure two-fermion problem
(but not in the two-fermion plus potential problem or in the
three-fermion problem). Other constraints could also be chosen, such as
that of Gross
\cite{3}, which puts one particle (normally the heaviest) on its mass shell.
We made a  nonexhaustive review in ref. \cite{7}.\par If we consider the
contributions of the poles of 
\begin{equation}G^0={1\over {1\over 2}P_0+p_0-h_1+i\epsilon h_1}
\,{1\over {1\over 2}P_0-p_0-h_2+i\epsilon h_2}\beta_1\beta_2 
\label{15}\end{equation} in an expression like $\,KG^0K,\,$ we must
perform an integration with respect to $\,p_0.\,$ If $\,K\,$ is
instantaneous, we get
\begin{equation}\int dp_0 G^0(p_0)\,=\,-2i \pi\tau\, g^0\,  
\beta_1 \beta_2,\qquad g^0\,=\,{1 \over
P_0-S+i\epsilon P_0 }  \label{18}\end{equation} where
\begin{equation}\tau={1\over 2} (\tau_1 + \tau_2), \qquad \tau_i = {h_i
\over \sqrt{h_i^2}} = {h_i \over E_i} = {\rm sign} (h_i). 
\label{19}\end{equation}  can also be written
\begin{equation}\tau=\Lambda^{++}-\Lambda^{--}, \qquad
\Lambda^{ij}=\Lambda_1^i\Lambda_2^j,
\qquad \Lambda_i^\pm={E_i\pm h_i\over 2E_i}. \label{20}\end{equation}     
When $\,K\,$ is not instantaneous, we must add the contributions of its
singularities. Furthermore, in the residues of the poles of
$\,G^0\,$ we must take $\,K\,$ at $\,p_{10}\!=\!h_1\,$ or at
$\,p_{20}\!=\!h_2,\,$ according to the chosen integration path and to the
sign of $\,\tau.\,$ In order to preserve an explicit symmetry between
the two fermions, we shall use the constraint $\,\delta(p_0\! -\! s),\,$ to
be combined with $\,g^0:$    
\begin{equation}G^{\delta S}(p_0)=-2i \pi\,\tau\,\delta(p_0\! -\!s)\,g^0\,  
\beta_1 \beta_2.\label{16}\end{equation}
 The operator $\tau$ has a clear
meaning in the basis built with the free solutions: it is +1 for
$h_1,h_2>0$, -1 for
$h_1,h_2<0$ and zero when they have opposite signs. It comes from the
dependence of the $p_0$ integral on the signs of the $i\epsilon h_i$. \par
The choice $\,G^{\delta}\!=\!G^{\delta S}\,$ has two merits: a) It leads
directly to Salpeter's equation
without higher-order correction terms to the potential when the Bethe-Salpeter
kernel is instantaneous. b) In the two-fermion plus potential problem and
in the three-fermion problem, the operator $\,\tau\,$ prevents the mixing
of asymptotically free fermions with opposite energy signs, which is the
origin of the continuum dissolution problem (see section 4 below). In
order to reduce the number of terms at lowest order, we shall however
keep only the $\,\Lambda^{++}\,$ part of $\,\tau\,$ and choose
\begin{equation}G^{\delta}(p_0)=G^{\delta P}(p_0)=-2i
\pi\,\Lambda^{++}\,\delta(p_0\! -\!s)\,g^0\,  
\beta_1 \beta_2.\label{17}\end{equation}
We shall examine some other possible choices in section 4.\par   
Let us now write the free propagator as the sum of the zero-order
propagator, plus a remainder:
\begin{equation}G^0=  G^{\delta}+G^R.  \label{23}\end{equation}  The
Bethe-Salpeter equation  becomes then the inhomogeneous equation
\begin{equation}\Phi=G^0K\Phi=(G^\delta +G^R)K\Phi=\Psi +G^RK\Phi,
\label{24}\end{equation}  with
\begin{equation}\Psi=G^\delta K\Phi \qquad (=G^\delta (G^0)^{-1}\Phi).
\label{25}\end{equation}  Solving (formally) the inhomogeneous equation
(\ref{24})  with respect to $\,\Phi\,$ and putting the result into
(\ref{25}), we get
\begin{equation}\Psi=G^\delta K(1-G^RK)^{-1}\Psi=G^\delta K^T\Psi  
\label{26}\end{equation}  where
\begin{equation}K^T=K(1-G^RK)^{-1}=K+KG^RK+...=(1-KG^R)^{-1}K 
\label{27}\end{equation}  obeys
\begin{equation}K^T=K+KG^RK^T=K+K^TG^RK. \label{28}\end{equation}  The
reduction series (\ref{27}) re-introduces in fact the reducible graphs
into the Bethe-Salpeter kernel, but with $G^0$ replaced by $G^R$.
Equation (\ref{26}) is a 3D equivalent of the Bethe-Salpeter equation. \par
 The relative energy dependence of eq. (\ref{26}) can be easily
eliminated:
\begin{equation}\Psi=\delta(p_0\! -\! s)\,\psi \label{29}\end{equation} 
and $\,\psi\,$ obeys:
\begin{equation}\psi\,=-2i\pi\,g^0\, \Lambda^{++} \,\int dp_0' dp_0
\delta(p'_0\! -\!s)\beta_1\beta_2K^T(p_0',p_0)\delta(p_0\!
-\!s)\,\psi.\label{32a}\end{equation}
Using the identity $\,\psi\!=\! \Lambda^{++} \psi,\,$ we can write 
\begin{equation}\psi\,=g^0\,V\,\psi,\qquad
V\,=-2i\pi\,\beta_1\beta_2K^{T++}(s,s),\label{30}\end{equation} 
\begin{equation}\beta_1\beta_2K^{T++}(s,s)\,\equiv\, \Lambda^{++} \!\int
dp_0' dp_0\delta(p'_0\! -\!s)\beta_1\beta_2K^T(p_0',p_0)\delta(p_0\!
-\!s)\, \Lambda^{++} .\label{32}\end{equation} Note that we write
$\,(p'_0,p_0)\,$ but
$\,(s,s),\,$ as we keep $\,s\,$ in operator form. This operator can be
diagonalized in the spatial momentum space by using the
$\,\Lambda^{ij}\,$ projectors. The eigenvalue will depend on the position
of $\,s\,$ in the formula: the eigenvalue of the first $\,s\,$ in
(\ref{32}) will be built with the final momenta and that of the last
$\,s\,$ will be built with the initial momenta.  \par
The inversion of the reduction is given by
\begin{equation}\Phi\,=\,(\,1-G^RK\,)^{-1}\,\Psi\,=\,
(\,1+G^RK^T\,)\,\Psi\,=\,(\,1+G^0K^T-G^{\,\delta}
K^T\,)\,\Psi\,=\,G^0K^T\,\Psi\label{32a*}\end{equation}
or, explicitating the relative energy:
\begin{equation}\Phi(p'_0)\,=\,G^0(p'_0)\,K^T(p'_0,s)\,\psi.
\label{32b}\end{equation}   
\par    The splitting of $\,G^0\,$ into two terms containing a
$\,\delta\,$ is the origin of unphysical singularities in the terms of
$\,K^T\,$ when the argument of the delta vanishes on the singularities of
$\,K.\,$ When the full $\,K^T\,$ is computed, the singularities of the
different terms cancel mutually. When $\,K^T\,$ is truncated, some of the
unphysical singularities have to be removed by hand \cite{3,12}.\par
This 3D reduction
can also be described in terms of transition operators. The 4D transition
operator is
\begin{equation}T\,=\,K\,+\,K\,G^0\,K\,+\,\cdots\label{*21}\end{equation}
\noindent and $\,K^T\,$ can be obtained by keeping only the
$\,G^R\,$ part of $\,G^0\,$ in it. We have also
$$T=K(1\!-\!G^0K)^{-1}=K(1\!-\!G^RK\!-\!G^{\delta}K)^{-1}=K(1\!-\!G^RK)^{-1}
(1\!-\!G^{\delta}K(1\!-\!G^RK)^{-1})^{-1}$$
\begin{equation}=K^T(1\!-\!G^{\delta}K^T)^{-1}=K^T+K^TG^{\delta}K^T+\cdots
\label{*22}\end{equation} 
 so that the 3D transition operator 
\begin{equation}T^{3D}\,=\,V\,+\,V\,g^0\,V\,+\,\cdots\label{*23}\end{equation}
is also given by
\begin{equation}T^{3D}\,=\,-2i\pi\,\beta_1\beta_2\,T^{++}(s,s).\label{*24}
\end{equation}
 We see that the 3D transition
operator is a constrained form of that of field theory. Both operators become
equal to the physical scattering amplitude when both fermions are on their
positive-energy mass shells. This was not
guaranteed a priori, as our original two-fermion Bethe-Salpeter equation
(\ref{1})  was valid only for bound states. Our 3D equation (\ref{30}) is
a bound state equation too. To include the scattering states we should
add an inhomogeneous term, or write the equation in the form
\begin{equation}(P_0-E_1-E_2\,)\,\psi\,=\,V\,\psi.\label{38.1}\end{equation}   
Conversely:
$$T=K^T+K^TG^{\delta}K^T+K^TG^{\delta}(K^T+K^TG^{\delta}K^T+\cdots)G^{\delta}
K^T$$
\begin{equation}=K^T+K^TG^{\delta}K^T+K^TG^{\delta}\,{-1\over2i\pi}\,
\beta_1\beta_2\,T^{3D}G^{\delta}K^T.
\label{*26b}
\end{equation}  A Bethe-Salpeter equation leading directly to the same 3D
reduction can be obtained by replacing  the kernel $\,K\,$ by the instantaneous
positive-energy kernel
$\,K^{T++}(s,s).\,$ In this case the 4D transition operator $\,T\,$ becomes
equal to
$\,T^{++}(s,s)=-\beta_1\beta_2\,T^{3D}/2i\pi.$\par
In contrast with the textbook cases, the potentials deduced from
field theory are in general energy-dependent (although one starts often
with an energy-independent approximation). We shall thus get a set of
eigenvalues $\,W^\lambda(P_0)\,$ corresponding to mutually orthogonal
eigenstates 
$\,\psi^\lambda(P_0)\,$ of the hamiltonian. The physical energy spectrum will be
given by the solutions of the algebric equations
\begin{equation}P_0\,=\,W^{\lambda}\,(P_0)\label{41}\end{equation} and the
corresponding wave functions are no more orthogonal with the the usual
scalar product, as they are no more eigenstates of the same hamiltonian.
The usual methods of perturbation calculation are to be revisited
\cite{19}.\par
\noindent The first-order correction term
to the energy is
\begin{equation}
P_0-P_0^{(0)}\,=\,\,-2i\pi\,<\Lambda^{++}\,\beta_1\beta_2\,(KG^RK)(s,s)\,
\Lambda^{++}>.
\label{*27}\end{equation} Since
$\,P_0^{(0)},\,\,(KG^RK)(s,s)\,$ and the zero-order wave functions are in
general total energy dependent,  (\ref{*27}) is in fact a numerical
equation in $\,P_0\,$ (or a matricial equation in case of degeneracy). If
the $\,P_0\,$ dependence of the zero-order potential is of higher-order, one
has only to take the
$\,P_0\,$ dependence of $\,P_0^{(0)}\,$ into account.\par
What is lost in our 3D equation compared with the starting Bethe-Salpeter
equation? The missing information is given by equation (\ref{32b}), which
gives the bound state solutions of the Bethe-Salpeter equation in terms of
the corresponding solutions of the 3D equation by adding the dependence in
the relative energy. Furthermore, the transition operators are the same on
the positive-energy mass shells, although the homogeneous Bethe-Salpeter
equation was originally deduced from the inhomogeneous Bethe-Salpeter
equation (for the propagators) by postulating a bound state \cite{1}. Were
some bound states lost in the 3D reduction, like the "abnormal solutions"
related to excitations in the relative time degree of freedom? We have
studied this still open problem elsewhere. Our conclusions were: 1). The
abnormal solutions of the Bethe-Salpeter equation, when they exist, can
also be obtained in the 3D equation via the total energy dependence of the
potential, leading to multiple solutions of (\ref{41}) for each
value of $\,\lambda\,$ \cite{a1,a2}. 2). The abnormal solutions found until now
are artefacts due to the ladder approximation, and should not appear when all
crossed graphs are included \cite{a2,a3}.     

\section{The three-fermion problem.}

\subsection{A cluster-separable three-fermion 3D equation.}
In the three-fermion framework, we shall append indices (ij)=(12),(23),(31)
to the two-fermion symbols and write $\,P_{ij},K_{ij},\,$ etc.... The index
k=1,2,3 will denote the third fermion. Some of these symbols will also get a
new three-fermion meaning when left unindexed, such as 
\begin{equation}P=p_1+p_2+p_3,\qquad
S=h_1+h_2+h_3.\label{41**}\end{equation}
A cluster-separable three-fermion 3D equation can be obtained by combining
three two-fermion potentials (\ref{30}):
\begin{equation}\psi={1\over
P_0-S}\,(\,V_{12}+V_{23}+V_{31})\psi\label{42**}\end{equation}
given by our 3D reduction in the two-fermion framework. The potential
$\,V\,$ given by (\ref{30}) depended on the total two-fermion energy
$\,P_0\,$ which we did not wrote as an argument, because it was conserved. 
When
using these two-fermion potentials in the three-fermion framework, we shall
replace the now nonconserved two-fermion energy argument $\,P_{ij0}\,$ by the
operator
$\,P_{ij0}\!+\!(p_{k0}\!-\!h_k)\!=\!P_0\!-\!h_k,\,$ which is the difference
between the conserved three-body energy and the free energy of the third
fermion. The potential $\,V_{ij}\,$ is thus no more strictly a two-fermion
term, and the third fermion is no more strictly a spectator when there are
interactions between the cluster (ij) and the fermion k.  At the (ij)+k
cluster-separated limit, however, the fermion k will be free and
$\,P_0\!-\!h_k\,$ will be replaced in $\,V_{ij}\,$ by the now conserved
$\,P_{ij0}.\,$     Let us indeed "switch off" the potentials
$\,V^0_{23}\,$ and
$\,V^0_{31}\,$ in equation (\ref{42**}). This equation can then be splitted
into
\begin{equation}\psi=\psi_{12}\psi_3,
\qquad P_0=P_{120}+p_{30},\label{59*}\end{equation}
\begin{equation}\psi_{12}={1\over P_{120}-E_1-E_2 
}V_{12}(P_{120})\,\psi_{12}, \qquad
(p_{30}-h_3)\,\psi_3=0,\label{60*}\end{equation}             
i.e. equation (\ref{30}) for the (12)
subsystem and a free Dirac equation for fermion 3.
Our 3D equation (\ref{42**}) satisfies thus clearly the cluster
separability requirement. Furthermore, the three cluster-separated limits
are exact equivalents of the corresponding two-fermion Bethe-Salpeter
equations. This cluster separability is a property of the equation, or,
more exactly, of the full Green function. For a given scattering solution
it is also possible to take the cluster-separated limit at fixed $\,P.\,$
This is not possible for the bound state solutions. \par
In replacing $\,P_{ij0}\,$ by $\,P_0\!-\!h_k\,$ we exploit the fact that
the opposite cluster is at the cluster-separated limit a free fermion. With
four fermions, we should introduce, for exemple, a potential $\,V_{34}\,$ 
inside a potential $\,V_{12}\,$ and vice-versa, getting an
unmanageable infinite nesting of potentials. This limits our recipe to the case
of three fermions.\par
  Our two
and three-fermion equations were written until now in an unspecified reference
frame. We shall consider three possible answers to the covariance
challenge:\par -- Write explicitly covariant equations. This is not the case
ou our two and three-fermion 3D equations, built around non-covariant
"approached" free propagators. We shall briefly present in section 4 a 3D
reduction method based on Sazdjian's covariant "approached" free propagator
\cite{18,7,14}.\par -- Choose the total rest frame $\,\vec P\!=\!\vec 0\,$ as
the reference frame. All quantities can then be made invariant by building
invariants with the 4-vector $\,P.$\par
-- Search for an implicit covariance by equivalence with an explicitly
covariant equation. A 3D equation obtained by reduction of an explicitly
covariant Bethe-Salpeter equation will for example be implicitly
covariant if all terms of the series generated by the reduction are
kept, even if the terms of the series are not separately covariant. If
the series is truncated, the 3D equation will be (implicitly) covariant
up to a given order only.\par
What about our equation (\ref{42**})? We can write it in the three-fermion
reference frame. If we "switch off" the potentials $\,V_{23}\,$ and
$\,V_{31},\,$ the equation for the subsystem (12) will still be written
in the global three-fermion rest frame, but this link with fermion 3 will
disappear if we go back to the original two-fermion Bethe-Salpeter
equation, which is explicitly covariant. If the series generated by the
reduction of this equation is truncated, we loose the exact covariance of
the two-fermion equation and the exact cluster separability of the
three-fermion equation.\par
A well known difficulty in the writing of equations for two
mutually interacting relativistic particles in an external potential and for
three or more mutually interacting relativistic particles is the
continuum dissolution problem: if we combine asymptotically free
particles with opposite energy signs, it becomes possible to build a
continuum of solutions at any given total energy. This renders the
building of normalisable bound state solutions impossible. The
introduction of our positive free-energy projectors is the usual way of
solving this problem. Our 3D equation has solutions built entierely in the
$\,\Lambda^{+++}\!=\!+1\,$ subspace and satisfying
\begin{equation}\psi={\Lambda^{+++} \over
P_0-E_1-E_2-E_3}\,(\,V_{12}+V_{23}+V_{31})\psi.\label{43**}\end{equation}
 We shall come back to this
point in section 4.\par

\subsection{Positive-energies instantaneous appproximation of the
three-fermion Bethe-Salpeter equation.}
 The Bethe-Salpeter equation for three fermions is
\begin{equation}\Phi=\left[G^0_1G^0_2K_{12}+G^0_2G^0_3K_{23}+G^0_3G^0_1K_{31}
+G^0_1G^0_2G^0_3K_{123}\right]
\Phi\label{*52}\end{equation}  where $K_{123}$ is the sum of the purely
three-body irreducible contributions. Let us first consider the case of three
two-body positive-energy, instantaneous and cluster-energy independent
interactions $\,K^0_{ij}\,$ (i.e. containing
$\,\Lambda^{++}_{ij}\,$ projectors  and independent of
$\,p'_{ij0},p_{ij0},P_{ij0}\,$).  The Bethe-Salpeter equation becomes in this
case
\begin{equation}\Phi=\,{-1\over2i\pi}\,G^0_1G^0_2G^0_3\,\beta_1\beta_2\beta_3\,
\left[\,V^0_{12}\,\psi_{12}\,+\,V^0_{23}
\,\psi_{23}\,+\,V^0_{31}\,\psi_{31}\,\right]\label{*53}\end{equation}
\begin{equation}V^0_{ij}\,=\,-2i\pi\,\beta_i\beta_jK^0_{ij},\qquad
\psi_{ij}(p_{k0})=\beta_k\,G_{0k}^{-1}\int
dp_{ij0}\,\Phi.\label{*54}\end{equation} 
Replacing $\,\Phi\,$ in (\ref{*54}) by its expression (\ref{*53})    leads to
a set of three coupled integral equations in the $\,\psi_{ij}:$
\begin{equation}\psi_{12}(p_{30})={-1\over2i\pi}\int
dp_{120}\,G^0_1G^0_2\,\beta_1\beta_2\left[\,
V^0_{12}\,\psi_{12}(p_{30})\,+\,
V^0_{23}\,\psi_{23}(p_{10})\,+\,V^0_{31}\,
\psi_{31}(p_{20})\,\right] \label{58k}\end{equation}
where $\,p_{10},p_{20}\,$ must be written in terms of
$\,P_0,p_{30},p_{120}:$   
\begin{equation}p_{10}\,=\,{P_0-p_{30}\over2}\,+\,p_{120},\qquad
p_{20}={P_0-p_{30}\over2}\,-\,p_{120}.
\label{59}\end{equation} and similarly for $\,\psi_{23}\,$ and
$\,\psi_{31}.\,$  We shall now search for solutions
$\,\psi_{ij}(p_{k0})\,$   
 analytical in the lower Im$(p_{k0})\!<\!0\,$ half plane and  perform the
integration (\ref{58k})
 by closing the integration paths around these regions.   The only
singularities will then be the poles of the free propagators. The result
is
\begin{equation}\psi_{12}(p_{30})=\,{\Lambda^{++}_{12}\over
(P_0-S)-(p_{30}-h_3)+i\epsilon}\,\left[\,V^0_{12}\psi_{12}(p_{30})
+V^0_{23}\psi_{23}(E_1)+V^0_{31}\psi_{31}(E_2)\,\right]\label{*55}\end{equation}
and similarly for $\,\psi_{23}\,$ and
$\,\psi_{31}\,$ (from now on, we shall often write the equations for the (12)
pair only, understanding the two other equations obtained by circular
permutations on the $\,ijk\,$ indexes). Solving (\ref{*55}) with respect to
$\,\psi_{12}(p_{30})\,$ gives a wave function which is analytical in the
Im($p_{30})\!<\!0\,$ half plane, and confirms thus the existence of such
solutions. Furthermore, equation (\ref{*55}) shows that the three projections
$\,\Lambda^+_k\psi_{ij}(E_k)\,$ are equal  (let us call them
$\,\psi\,$) and satisfy the 3D equation
\begin{equation}\psi={\Lambda^{+++}\over
P_0-E_1-E_2-E_3}\,\left[\,V^0_{12}+V^0_{23}+V^0_{31}\,\right]\,\psi.\label{*56}
\end{equation}
Moreover,  $\,\psi\,$ is the integral of the Bethe-Salpeter amplitude with
respect to the relative times:         
\begin{equation}\psi={-1\over2i\pi}\,\int dp_0\,\Phi\equiv{-1\over2i\pi}\,\int
dp_{10}dp_{20}dp_{30}\,\delta(p_{10}+p_{20}+p_{30}-P_0)\,\Phi.\label{*57}
\end{equation}
Let us now choose
\begin{equation}K^0_{ij}\,=\,
K^{T++}_{ij}(\,s_{ij},\,s_{ij},\,P_0-h_k\,)\label{*58}
\end{equation}
The 3D equation (\ref{*56}) is then built with the two-body potentials
$\,V^0_{ij}\!=\!V_{ij}(P_0\!-\!h_k),\,$ with the $\,V_{ij}\,$ defined by
(\ref{30}) in section 2 for each fermion pair, and 
becomes equation (\ref{43**}). \par
 Let us now consider a non-approximated
three-fermion Bethe-Salpeter equation (\ref{*52}) and try to transform it
into our 3D equation (\ref{*56}) above, completed by higher-order correction
terms. It first seemed us a good idea to start with approximations giving the
3D equation (\ref{*56}) in a more direct way than above, like
\begin{equation}G^0_1G^0_2\,\approx\,G^{\delta P}_{12}\label{*59}\end{equation}
\begin{equation}G^0_1G^0_2\,\delta(p'_{30}-p_{30})\,\approx\,G^{\delta P}_{12}\,
\delta(p'_{30}-h_3)\label{*60}\end{equation}
\begin{equation}K_{12}\,\delta(p'_{30}-p_{30})\,\approx\,\Lambda^+_3\,
K^0_{12}\,\delta(p'_{30}-E_3).\label{*61}\end{equation} 
Each of these
approximations gives in each case the 3D equation (\ref{*56}), but the
correction terms spoil the result: with (\ref{*59}) or (\ref{*60}) they contain
supplementary zero-order contributions. With (\ref{*61}) they contain no
supplementary zero-order contributions, but the correction terms can not easily
be sorted by increasing order. We must thus proceed more carefully. We finally
decided to start with positive-energy, instantaneous and cluster-energy
independent approximations of the two-body kernels, like (\ref{*58}) or its Born
approximation, adding the third
$\,\Lambda^+_k\,$ projector in order to avoid unnecessary complications, and to
proceed in three steps, inspired by the demonstration
(\ref{*52})-(\ref{*56}):\par i) Transform the Bethe-Salpeter equation 
(\ref{*52}) into a set of three coupled equations for the
$\,\psi_{ij}(p_{k0}).$\par ii) Transform this set into a set of three coupled
3D equations for the $\,\psi_{ij}(E_k)\,$ (these are {\it not} the Faddeev
equations).\par iii) Transform this last set into a single 3D equation for
$\,\psi,\,$ defined as the mean of the
$\,\psi_{ij}(E_k).$\par         

\subsection{Three coupled equations for three cluster-energy depending
amplitudes.}
Let us now consider the unapproximated Bethe-Salpeter equation (\ref{*52})
and define
\begin{equation}G^0\,=\,G^0_1\,G^0_2\,G^0_3\,,\label{*63}\end{equation}
\begin{equation}K\,=\,(G^0_3)^{-1}K_{12}\,+\,(G^0_1)^{-1}K_{23}\,+\,
(G^0_2)^{-1}K_{31}
\,+\,K_{123}.\label{*64}\end{equation}
The Bethe-Salpeter equation (\ref{*52}) takes then the compact form:
\begin{equation}\Phi\,=\,G^0K\,\Phi.\label{*62}\end{equation}
 Let us
also define
\begin{equation}K^0\,=\,(G^0_3)^{-1}K^0_{12}\,+\,(G^0_1)^{-1}K^0_{23}\,+\,
(G^0_2)^{-1}K^0_{31}\,,\label{*64b}\end{equation}
\begin{equation}K^R\,=\,K-K^0\,,\label{64c}\end{equation} 
each $\,K^0_{ij}\,$ being independent of all relative energies and containing
$\,\Lambda^{+++}\,$ projectors (in order to render the following
calculations simpler, we now add the $\,\Lambda^{+}_k\, $ projectors to the
$\,\Lambda^{++}_{ij}\, $ projectors of the previous section). 
Let us now perform the transformations
\begin{equation}\Phi=G^0K^0\Phi+G^0K^R\Phi\label{*66}\end{equation}  
\begin{equation}\Phi=(1-G^0K^R)^{-1}G^0K^0\Phi=G^K\,K^0\Phi\label{*67}
\end{equation}
where
\begin{equation}G^K\,=\,G^0+G^0K^RG^0+G^0K^RG^0K^RG^0+\cdots\equiv\,G^0+G^{KR}.
\label{*68}
\end{equation} Since eq. (\ref{*67}) ends with $\,K^0\Phi,\,$ we can transform
it into a set of three coupled equations for the
$\,\psi_{ij}(p_{k0}),\,$ now defined with  the $\,\Lambda^{+++}\,$ projector:
\begin{equation}
\psi_{ij}(p_{k0})=\Lambda^{+++}\,\beta_k\,G_{0k}^{-1}\int
dp_{ij0}\,\Phi.\label{*68b}\end{equation}
We get
\begin{equation}\psi_{12}(p'_{30})\,=\,\sum_k\int
dp_{k0}\,G^K_{12,ij}(p'_{30},p_{k0})\,K^0_{ij}\,\psi_{ij}(p_{k0})\label{*69}
\end{equation} 
with
\begin{equation}G^K_{12,ij}(p'_{30},p_{k0})\,=\,\Lambda^{+++}\,\beta_3\,
\big[G^0_3(p'_{30})
\big]^{-1}\int
dp'_{120}\,dp_{ij0}\,G^K(p'_0,p_0)\,\,\beta_k\label{*70}\end{equation}

\subsection{Three coupled 3D equations.} Let us now compute the  contribution
of $\,G^0\,$ to (\ref{*70}). We get (omitting the $\,\Lambda^{+++}\,$ already
contained in $\,K^0$):  
\begin{equation}G^0_{12,12}(p'_{30},p_{30})\,=\,{-2i\pi\over
(P_0-S)-(p'_{30}-E_3)+i\epsilon}\,
\beta_1\beta_2\,\delta(p'_{30}-p_{30})\label{*72}\end{equation} 
\begin{equation}G^0_{12,23}(p'_{30},p_{10})\,=\,\beta_3\,G^0_1(p_{10})
\,G^0_2(P_0-p_{10}
-p'_{30})\,\beta_1\label{*73}\end{equation}   
\begin{equation}G^0_{12,31}(p'_{30},p_{20})\,=\,\beta_3\,G^0_2(p_{20})
\,G^0_1(P_0-p_{20}
-p'_{30})\,\beta_2\label{*74}\end{equation} The product of propagators
(\ref{*73})  will be included in a clockwise path integral in
$\,p_{10}.\,$ With positive-energy instantaneous kernels, this integral is given
by the residue of the pole of
$\,G^0_1(p_{10})\,$ at
$\,p_{10}\!=\!E_1.\,$  This is no more true in the general case, but
we can still isolate a part of $\,G^0_1(p_{10})\,$ which leads to the same 
result:
\begin{equation}G^0_1(p_{10})\,=\,G^{\delta}_1(p_{10})\,+\,G^R_1(p_{10})
\label{*75}\end{equation}
with
\begin{equation}G^{\delta}_1(p_{10})\,=\,-2i\pi\,\delta(p_{10}-E_1)\,\beta_1
\label{*76}\end{equation}
\begin{equation}G^R_1(p_{10})\,=\,{1\over
p_{10}-E_1-i\epsilon}\,\,\beta_1\label{*77}\end{equation} so that
\begin{equation}G^{\delta}_{12,23}(p'_{30},p_{10})\,=\,{-2i\pi\over
(P_0-S)-(p'_{30}-E_3)+i\epsilon}\,
\beta_2\beta_3\,\delta(p_{10}-E_1)\label{*78}\end{equation} and similably 
\begin{equation}G^{\delta}_{12,31}(p'_{30},p_{20})\,=\,{-2i\pi\over
(P_0-S)-(p'_{30}-E_3)+i\epsilon}\,
\beta_3\beta_1\,\delta(p_{20}-E_2).\label{*79}\end{equation} Putting all
together, we have finally
$$\psi_{12}(p'_{30})\,=\,{-2i\pi\over (P_0-S)-(p'_{30}-E_3)+i\epsilon}\,
\beta_1\beta_2\,K^0_{12}\,\psi_{12}(p'_{30})$$ 
$$+\,{-2i\pi\over (P_0-S)-(p'_{30}-E_3)+i\epsilon}\,
\big[\,\beta_2\beta_3\,K^0_{23}\,\psi_{23}(E_1)\,+\,\beta_3\beta_1\,K^0_{31}\,
\psi_{31}(E_2)\,\big]$$          
$$+\,\int
dp_{10}\,\beta_3\,G^R_1(p_{10})\,G^0_2(P_0-p_{10}-p'_{30})\,\beta_1\,K^0_{23}\,
\psi_{23}(p_{10})$$
$$+\,\int
dp_{20}\,\beta_3\,G^R_2(p_{20})\,G^0_1(P_0-p_{20}-p'_{30})\,\beta_2\,K^0_{31}\,
\psi_{31}(p_{20})$$
\begin{equation}+\,\sum_k\int
dp_{k0}\,G^{KR}_{12,ij}(p'_{30},p_{k0})\,K^0_{ij}\,\psi_{ij}(p_{k0})
\label{*80}\end{equation}
If we neglect the last term of (\ref{*80}) the two preceding terms vanish also by
integration, and we recover the result of the positive-energy instantaneous
approximation. We shall thus move the first term of (\ref{*80}) to the left-hand
side, consider the two next terms (second line) as the principal contributions,
and the three last terms as perturbations. The difference between this approach
and that based on (\ref{*61}) lies in the fact that we do not replace
immediately $\,p'_{30}\,$ by
$\,E_3\,$ in the first term. We get
$$\psi_{12}(p'_{30})\,=\,g_{12}(p'_{30})\,\big[\,V^0_{23}\,\psi_{23}(E_1)\,+
\,V^0_{31}\,
\psi_{31}(E_2)\,\big]$$
\begin{equation}+\,\,g_{12}(p'_{30})\,[\,g^0_{12}(p'_{30})\,]^{-1}\,\sum_k\int
dp_{k0}\,G^{KT}_{12,ij}(p'_{30},p_{k0})\,K^0_{ij}\,\psi_{ij}(p_{k0})
\label{*81}\end{equation}
where $\,G^{KT}_{12,ij}\,$ is defined from $\,G^{KR}_{12,ij}\,$ by including
the 3th and the 4th lines of (\ref{*80}) in the last one, and 
\begin{equation}g_{12}(p'_{30})\,=\,{1\over
(P_0-S-V^0_{12})-(p'_{30}-E_3)+i\epsilon}\label{*82}\end{equation}
\begin{equation}g^0_{12}(p'_{30})\,=\,{1\over
(P_0-S)-(p'_{30}-E_3)+i\epsilon}\label{*83}\end{equation}  
 The iterations of (\ref{*81}) lead to
$$\psi_{12}(p'_{30})\,=\,g_{12}(p'_{30})\,\big[\,V^0_{23}\,\psi_{23}(E_1)\,+
\,V^0_{31}\,
\psi_{31}(E_2)\,\big]$$
$$+\,\,g_{12}(p'_{30})\,[\,g^0_{12}(p'_{30})\,]^{-1}\,\sum_k\int
dp_{k0}\,G^{\,TT}_{12,ij}(p'_{30},p_{k0})\,K^0_{ij}$$
\begin{equation}g_{ij}(p_{k0})\,\big[\,V^0_{jk}\,\psi_{jk}(E_i)\,+\,V^0_{ki}\,
\psi_{ki}(E_j)\,
\big].\label{*84}\end{equation} defining $\,G^{TT}\,$ as the sum of the
iterations of
$\,G^{KT}:$
$$G^{\,TT}_{12,ij}(p'_{30},p_{k0})\,=\,G^{\,KT}_{12,ij}(p'_{30},p_{k0})\,+\,
\sum_{k'}\int
dp''_{k'0}\,G^{\,KT}_{12,i'j'}(p'_{30},p''_{k'0})\,K^0_{i'j'}$$
\begin{equation}
g_{i'j'}(p''_{k'0})\,[\,g^0_{i'j'}(p''_{k'0})\,]^{-1}
G^{\,KT}_{i'j',ij}(p''_{k'0},p_{k0})\,+\cdots\label{*84.1}\end{equation}
\hfill\break\noindent  
 We can now take (\ref{*84}) at
$\,p'_{30}\!=\!E_3,\,$ multiply at left by
$\,[\,g_{12}(E_3)]^{-1},\,$  and move the term in
$\,V^0_{12}\psi_{12}(E_3)\,$ back to the right-hand side:   
$$\psi_{12}(E_3)\,=\,{1\over
P_0-S+i\epsilon}\,\big[\,V^0_{12}\,\psi_{12}(E_3)\,+\,V^0_{23}\,
\psi_{23}(E_1)\,+\,V^0_{31}\,
\psi_{31}(E_2)\,\big]$$
\begin{equation}+\,\sum_k\,M_{12,ij}\,\big[\,V^0_{jk}\,\psi_{jk}(E_i)\,+
\,V^0_{ki}\,
\psi_{ki}(E_j)\,\big]\label{*85}\end{equation} with
\begin{equation}M_{12,ij}\,=\,\int
dp_{k0}\,G^{\,TT}_{12,ij}(E_3,p_{k0})\,K^0_{ij}\,g_{ij}(p_{k0}).\label{*86}
\end{equation} 

\subsection{One 3D equation.} Let us write (\ref{*85}) and the two other
equations obtained by circular permutations in matricial form. We shall define
\begin{equation}\Psi^k\,=\,\psi_{ij}(E_k),\qquad
\Pi^{ij}\,=\,{1\over3},\quad \overline\Pi
\,=\,1-\Pi.\label{*87}\end{equation} The matrix $\,\Pi\,$ is the projector
($\Pi^2\!=\!\Pi\,$) on the $\,u^k\!=\!1\,$ vector (of norm 3). We have
\begin{equation}\Pi\,\Psi\,=\,\psi\,u\,,\qquad
\psi\,=\,{1\over3}\,(\,\Psi^1\,+\,\Psi^2\,+\,\Psi^3\,).\label{*88}\end{equation}
We can write eq. (\ref{*85}) in the form
\begin{equation}
\Psi\,=\,M\,\Psi,\quad M=M^0+M^R,\quad M^{0i1}\,=\,{1\over P_0-S+i\epsilon}\,
V^0_{23},\,\, etc...\label{*89}\end{equation} The action of the projector
$\,\Pi\,$ on
$\,M^0\,$ is
\begin{equation}\Pi\,M^0\,=\,M^0,\quad M^0\,\Pi\,=\,{\Pi\over
P_0-S+i\epsilon}\,V^0,\quad V=V^0_{12}+V^0_{23}+V^0_{31}.\label{*90}
\end{equation}
We shall now transform the matricial equation (\ref{*89}) into a scalar equation
for $\,\psi.\,$ The expansion of (\ref{*89}) with respect to
$\,\Pi\Psi\,$ gives
\begin{equation}\Pi\,\Psi\,=\,\Pi\,M\,(\,1+\overline\Pi M+\overline\Pi
M\,\overline\Pi M+\cdots)\,\Pi\,\Psi\label{*91}\end{equation} in which
$\,\overline\Pi M\!=\!\overline\Pi M^R.\,$ Taking the scalar product with
$\,u\,$ leads finally to
\begin{equation}\psi\,=\,\left[\,{1\over
P_0-S+i\epsilon}\,V^0\,+\,{1\over3}\,u^{\top}\,(\,M^R+M\,[\,\overline\Pi
M^R+\overline\Pi M^R\,\overline\Pi
M^R+\cdots]\,)\,u\,\right]\,\psi.\label{*92}\end{equation} In this equation,
$\,V^0\,$ contains the projector $\,\Lambda^{+++}\,$ at left and at right, while
$\,M^R\,$ contains it at right. We can thus trivially transform (\ref{*92}) into
an equation for
$\,\Lambda^{+++}\psi.\,$            
             
\subsection{First-order energy shift.} The 3D reduction above is rather
complicated, since it implies three levels of series expansions. The
computation of the first-order energy shift remains however tractable. The
energy shift is, keeping only the terms which contribute at first-order:
\begin{equation}P_0-P_0^{(0)}\,=\,<(P_0-S)\,{1\over3}\,u^{\top}\,(\,1+M^0
\overline\Pi\,)\,
M^R\,u>.\label{*93}\end{equation} We can make the replacements
\begin{equation}1+M^0\overline\Pi\,=\,1+M^0-{\Pi\over
P_0-S}\,V^0\,\to\,M^0\label{*94}\end{equation} as the contributions of the first
and third terms cancel mutually. We remain with
$$P_0-P_0^{(0)}\,=\,<(P_0-S)\,{1\over3}\,u^{\top}\,M^0\,M^R\,u>$$
\begin{equation}=\,\sum_{k',k}<V^0_{i'j'}\,M_{i'j',ij}\,(\,V^0-V^0_{ij}\,)>
\label{*95}\end{equation} 
The contribution of the terms with $\,i'j'\!=\!12\,$ is then
\begin{equation}\Delta_{12}\,=\,<V^0_{12}\,\sum_k\int
dp_{k0}\,G^{\,TT}_{12,ij}(E_3,p_{k0})\,K^0_{ij}\,g_{ij}(p_{k0})\,
(\,V^0-V^0_{ij}\,)>
\label{*96}\end{equation} At first order, we can replace $\,G^{TT}_{12,ij}\,$ by
the non-iterated
$\,G^{KT}_{12,ij}\,$ and further by
$\,G^{KR}_{12,ij},\,$ as the 3th and 4th lines of (\ref{*80}) will lead to null
integrals. Using the definitions (\ref{*68}), (\ref{*70}) we get
$$\Delta_{12}\,=\,<V^0_{12}\,\sum_k\int\delta(p'_{30}-E_3)
dp'_0\,dp_0\,\beta_3\,\big[G^0_3(p'_{30})\big]^{-1}$$
\begin{equation}G^{KR}(p'_0,p_0)\,\,\beta_k\,K^0_{ij}\,g_{ij}(p_{k0})\,
(\,V-V^0_{ij}\,)>
\label{*97}\end{equation}\vskip5mm\noindent Replacing $\,V\,$ by $\,P_0\!-\!S\,$
and $\,G^{KR}\,$ by
$\,G^0K^RG^0\,$ leads to
\begin{equation}P_0-P_0^{(0)}\,=\,{-1\over4\pi^2}<\int
dp'_0\,dp_0\,V_F(p'_0)\,G^0(p'_0)K^R(p'_0,p_0)G^0(p_0)\,\beta_1\beta_2
\beta_3\,V_I(p_0)>
\label{*98}\end{equation}          with
\begin{equation}V_I(p_0)\,=\,\sum_k\,V^0_{ij}
\,g_{ij}(p_{k0})\,g_{ij}^{-1}(E_k)\label{*99}\end{equation}
\begin{equation}V_F(p'_0)\,=-2i\pi\,\sum_{k'}\,V^0_{i'j'}\,
\delta(p'_{k'0}-E'_{k'})
\,\beta_{k'}\,\big[G^0_{k'}(p'_{k'0})\big]^{-1}.\label{*100}\end{equation} In
writing (\ref{*100}), we mean that
$\,\big[G^0_{k'}(p'_{k'0})\big]^{-1}$ cancels the 
$\,G^0_{k'}(p'_{k'0})\,$  contained in
$\,G^0(p'_0)\,$ before the replacement of $\,p'_{k'0}\,$ by
$\,E'_{k'}.\,$     The energy shift (\ref{*98}) contains 36 terms: there are 4
terms in
$\,K^R\,$ (if we do not forget the irreducible three-body term) times 3 terms
in $\,V_I\,$ and 3 terms in $\,V_F\,$.  Our reason for introducing these
$\,V_I\,$ and $\,V_F\,$ is the fact that they both could be replaced by
$\,V^0\,$  if
$\,K^R\,$ were instantaneous.\par The singularities to consider when performing
the integrations are the poles of the propagators, the singularities of
$\,K^R\,$ and the poles of $\,V_I.\,$ It is always possible to choose the
integration paths, in each term, in order to avoid contributions from the poles
of $\,V_I.\,$  The contributions of the poles of the propagators to the energy
shift  (\ref{*98}) are then easy to compute, as 
$\,g_{ij}(p_{k0})\,$ in (\ref{*99}) is to be taken at 
$\,p_{k0}\!=\!h_k.\,$ We get 
\begin{equation}R_{12}\,=\,<\sum_{k'k}\,V^0_{i'j'}\,{1\over
P_0-S}\,V^{Rba}_{12}\,{1\over P_0-S}\,V^0_{ij}>\label{*101}\end{equation} plus 
the similar
$\,R_{23}\,$ and $\,R_{31}\,$ and a contribution $\,R_{123}\,$ of the
irreducible three-body kernel.  By
$\,V^{Rba}_{12}\,$ we denote
$\,-2i\pi\beta_1\beta_2K^R_{12}\,$ with the initial fermion a, the final
fermion b  and the spectator fermion 3 on their positive-energy mass shells. We
have to consider 9 terms in 4 different groups:\hfill\break\par
$\,k'=k=3\,$\qquad\qquad\qquad  $\,ba=11,\,12,\,21\,$ or $\,22$\par
$\,k'=1\,\hbox{or}\,2,\,k=3\,$\qquad\qquad  $\,ba=k'1\,$ or
$\,k'2$\par
$\,k'=3,\,k=1\,\hbox{or}\,2\,$\qquad \qquad $\,ba=1\,k\,$ or
$\,2\,k$\par
$\,k'=1\,\hbox{or}\,2\,,\,k=1\,\hbox{or}\,2\,$\qquad\qquad 
$\,ba=k'k\,$\par\hfill\break We still have some freedom in the choice of
$\,ba\,$ which we could use to get a real energy shift, for example by choosing
$\,a\!=\!b\,$ in the three first groups. The contributions of an instantaneous
part
$\,K^{RI}_{12}\,$ of $\,K^{R}_{12}\,$ would sum into  a
$\,<V^{RI}_{12}>\,$ but we could also include this
$\,K^{RI}_{12}\,$ directly in $\,K^0_{12}.\,$ The corrections corresponding to
the differences between the various
$\,V^{Rba}_{12}\,$ are a step towards Gross' spectator model (see below,
section 6 and
\cite{12,13}).\par The principal task remains however the calculation of the
contributions of the singularities of the
$\,K^R_{ij},\,$ for the integration paths defined by our choices of
$\,ab\,$ above. As we shall see in section 5.3, these contributions could be
as important as these coming directly from the relative time dependence of
these kernels (they could however be partially cancelled by the contributions
of two-body non-ladder terms and three-body terms \cite{3,7,12} ). A
general expression of the contributions of these singularities can be
obtained by splitting the propagators containing the poles considered in
(\ref{*101}), using (\ref{*75}), the all-$G^\delta\,$ part giving
(\ref{*101}). \par For more explicit results, we should specialize to
specific three-fermion problems. It must be noted that the ratio
$\,g_{ij}(p_{k0})\,g_{ij}^{-1}(E_k)\,$ of (\ref{*99}) is no more 1 in the terms
containing
$\,G^R_k(p_{k0}).\,$ If the singularities of
$\,K^R\,$ in the complex $\,p_{k0}\,$ plane lie far from the positive-energy
mass shell, this ratio can be approximated by 
$\,-(p_{k0}\!-\!E_k)^{-1}\,g_{ij}^{-1}(E_k)\,$ and takes a relatively low value.
This suggests us to neglect the contributions of the terms containing
$\,G^R_k(p_{k0})\,$ to (\ref{*98}), as these contributions contain at least one
large denominator more than the contributions of the terms containing
$\,G^\delta_k(p_{k0})\,$. This leads to the replacement
\begin{equation}V_I(p_0)\to\,=-2i\pi\,\sum_{k\neq
u}\,V^0_{ij}\,\beta_k\,\big[G^0_k(p_{k0})\big]^{-1}\delta(p_{k0}-E_k)\,
+\,\sum_{k= u}\,V^0_{ij}\label{*104}\end{equation} in (\ref{*98}). We isolated the
contributions of the unconnected terms $\,k\!=\!u\,$ (like $\,k\!=\!3\,$
combined with
$\,k'\!=\!3\,$ and $\,K^R_{12}\,$). The resulting energy shift is now manifestly
real. 

\subsection{Cluster separability.} Physically, if we "switch off" the (23),
(31) and (123) interactions, we remain with a pair of interacting fermions
(12) and a free fermion (3). We would like to read this on our final 3D
equation. however:\par -- The cluster-separability condition may be satisfied
by an exact 3D reduction, but not by its truncated forms.\par -- Even with an
exact 3D reduction, the cluster separability property is a welcome bonus, not a
requirement: the Bethe-Salpeter equation (\ref{*52}), from which we started, is
indeed correct for three-fermion bound states only.\par In our calculations
above, the cluster-separability property is in general spoiled by our iteration
of (\ref{*81}), which starts with a term which vanishes at the (12)
cluster-separated limit. The (12)-limit  equation is then built with the
approached
$\,-2i\pi\beta_1\beta_2 K^0_{12}\,$ only. The cluster separability can however
be restored by performing a supplementary iteration, like our replacement of
$\,V^0\,$ by
$\,P_0\!-\!S\,$ in (\ref{*95}), leading to the cluster-separable correction
(\ref{*101}). Another possibility consists in our previous choice (\ref{*58})
  of the
$\,K^0_{ij}.\,$ With this choice, the cluster-separated limits are exact.
Moreover, the corresponding two-body scattering amplitudes are also correct
(of course, we do not mean that we should keep all the terms of the
$\,K^0_{ij}\,$ in an actual computation: we should only keep the ones which
contribute up to the desired order).   Simpler choices (as the Born
approximations) would however not be uncorrect in the calculation of the
three-fermion bound state spectrum.

\section{Other propagator-based 3D reductions.}
\subsection{Two fermions.}
We have already seen two different choices of $\,G^{\delta}:\,$ Salpeter's
propagator $\,G^{\delta S}\,$ (\ref{16}) and the positive-energy propagator
$\,G^{\delta P}\,$ (\ref{17}), which we used through this work until now. We
could also use Breit's propagator \cite{17}: 
\begin{equation}G^{\delta B}(p_0)=-2i \pi\,\delta(p_0\! -\!s)\,g^0\,  
\beta_1 \beta_2.\label{96*}\end{equation}
in the two-fermion problem only, as it would lead to continuum dissolution in
the two-fermion in an external potential and in the three-fermion problems.
The first-order constraint $\,\delta(p_0\!-\!s)\,$ could also be replaced by
the second-order constraint
$\,\delta(p_0\!-\!\mu),\,$ which leads to a simpler one-body limit, especially
with Breit's propagator (see section 7 below). In the spirit of
Gross'spectator model, we could also choose the unsymmetrical or
symmetrized propagators 
\begin{equation}G^{\delta G1}(p_0)=-2i \pi\,\Lambda^{+}_2\delta(p_{20}\!
-\!E_2)\,g^0\,  
\beta_1 \beta_2.\label{97*}\end{equation}
\begin{equation}G^{\delta GS}(p_0)=-2i
\pi\,{1\over2}\,\left[\Lambda^{+}_2\delta(p_{20}\!
-\!E_2)+\Lambda^{+}_1\delta(p_{10}\! -\!E_1)\right]\,g^0\,  
\beta_1 \beta_2.\label{98*}\end{equation}
in the two-fermion problem (for the three-fermion problem, see below
section 6).\par
With Sazdjian's covariant propagator \cite{18,7,14} 
$$G^{SZ}\,=\,-2i\pi\,\,{(p_1\cdot
\gamma_1+m_1)\,(p_2\cdot\gamma_2+m_2)\over
p_1^2+p_2^2-(m_1^2+m_2^2)+i\epsilon}\,\,\delta\,(P\cdot
p-{m_1^2-m_2^2\over2}\,)$$ 
\begin{equation}=-2i \pi\,{P_0+S\over2\vert
P_0\vert}\,\delta(p_0\! -\!\mu)\,g^0\,  
\beta_1 \beta_2.\label{99*}\end{equation}
the 3D potential is given by a series of separately covariant terms, so that
we can truncate it without breaking the covariance. In order to preserve
this property in the two-fermion in an external potential problem and in the
three-fermion problem, without meeting continuum dissolution, we must
however combine this propagator with a covariant substitute of the
noncovariant $\,\Lambda^{++}.\,$ A possible choice is the product
\begin{equation}\theta^{++}=\theta(p_{10})\theta(p_1^2)\theta(p_{20})
\theta(p_2^2)=\theta(p_{10}\!-\!\vert\,\vec
p_1\vert\,)\,\theta(p_{20}\!-\!\vert\,\vec
p_2\vert\,)\label{100*}\end{equation}
in which the constraint will replace $\,p_{i0}\,$ by
$\,{1\over2}P_0\pm\mu.\,$ The sign of $\,p_{i0}\,$ is indeed invariant
when $\,p_i^2>0.\,$ Another difficulty comes from the $\,{(P_0+S)/2\vert
P_0\vert}\,$ operator which obliges us to perform supplementary
manipulations after the 3D reduction in order to render the potential
hermitian, so that the Born term itself is finally given by a series.   We
examined elsewhere
\cite{20} how to use Sazdjian's propagator in the two-fermion, two fermions
in an external potential and three-fermion problems, leaving the reader decide
wether the covariance of the approximations balances the supplementary
complications.
\par With Salpeter's propagator, the 3D potential will be
\begin{equation}
V^S=-2i\pi\,\tau\,\beta_1\beta_2\,K^{TS}(s,s)\,\tau^2\label{101*}\end{equation}      
as $\,\tau^2\psi^{S}\!=\!\psi^S,\,$ with $\,K^{TS}\,$ built with
$\,G^{\delta S}\,$ like $\,K^{TP}\,$ with $\,G^{\delta P}.\,$ This potential
is hermitian in the $\,\tau^2\!=\!1\,$
subspace at fixed $\,P\,$ with the scalar product
\begin{equation}(\psi_i,\psi_j)\,=\,\int d^3p\,\,\psi^+_i(\vec
p\,)\,\tau(\vec p\,)\,\psi_j(\vec p\,).\label{33.1}\end{equation}

\subsection{Three fermions and continuum dissolution.}
The results of a 3D reduction of the two-fermion Bethe-Salpeter equation
can be used in the three-fermion problem at two levels: by choosing the
corresponding approximations of the two-fermion kernels in
the Bethe-Salpeter equation (\ref{*52}) , or by combining directly the
corresponding potentials into a 3D equation (\ref{42**}). Both approaches turn
out to be equivalent in the case of our 3D reduction based on $\,G^{\delta
P}.\,$ With the other choices, the 3D equation obtained by reducing the
approximated Bethe-Salpeter equation would contain supplementary terms. In
all cases, we get an exact three-fermion equation by "switching off" the
interactions with the third fermion. However, the three-fermion 3D equation
built with the potentials
$\,V^B$ or $\,V^{SZ},\,$ based on approximated propagators without
projectors, suffers of continuum dissolution.  It is  indeed
possible to build a continuum of solutions with any a priori given total
energy by combining asymptotically free fermions with opposite energy
signs. Any physical bound state is thus  mixed with such a continuum
and the building of normalisable bound state wave functions becomes
impossible [13-16]. In the pure two-body case the  energy of a
system
$\,(+,-)\,$ in the total rest frame is 
\begin{equation}E_1-E_2=\,\sqrt{\vec p^2+m_1^2}-\sqrt{\vec
p^2+m_2^2}\,=\,{m_1^2-m_2^2\over \sqrt{\vec p^2+m_1^2}+\sqrt{\vec
p^2+m_2^2}}\label{45}\end{equation} and lies thus between $\,m_1-m_2\,$
(whichever the sign) and zero.  We have thus no problem if  we make the
 assumption that the energies of the bound states lie between
$\,\vert\, m_1-m_2\vert\,$ and $\,m_1+m_2.$ Below $\,\vert\,
m_1-m_2\vert\,$ we meet the "strong field" problem. The strong field
problem for the one-body plus potential and two-body systems, the
continuum dissolution problem for the two-body plus potential and
three-body systems are both consequences of the possibility of pair
creation.\par  In the three-body case, the mixing with asymptotically
separated (12)(3) subsystems of opposite energy signs is excluded by the
total momentum conservation, as in the two-fermion case. For the mixing
with three-fermion asymptotically free states, let us consider for example
\begin{equation}E_3-E_1-E_2=\,\sqrt{\vec p_3^2+m_3^2}-\sqrt{\vec
p_1^2+m_1^2}-\sqrt{\vec p_2^2+m_2^2\,},\qquad\vec p_1+\vec p_2+\vec
p_3=\vec 0.\label{46}\end{equation}
 There is no lowest value (we can for example have $\,\vec p_3\!=\!0\,$
and $\,\vec p_1\!=\!-\vec p_2\,$ arbitrarily large).  The highest values
are obtained when $\,\vec p_1\,$ and $\,\vec p_2\,$ have the same
direction, and for $\,\vec p_1/m_1\!=\!\vec p_2/m_2\!=\!-\,\vec
p_3/(m_1\!+\!m_2).\,$ In this case we have
\begin{equation}E_3-E_1-E_2=\,\sqrt{\vec p_3^2+m_3^2}-\sqrt{\vec
p_3^2+(m_1+m_2)^2}\label{47}\end{equation} so that $\,E_3-E_1-E_2\,$ lies
finally between $\,-\infty\,$ and
$\,(m_3\!-\!m_1\!-\!m_2)\theta(m_3\!-\!m_1\!-\!m_2).\,$ Symmetrically,
$\,E_1+E_2-E_3\,$ lies between 
$\,(m_1\!+\!m_2\!-\!m_3)\theta(m_3\!-\!m_1\!-\!m_2)\,$ and $\,\infty.\,$
If we assume that the energies of the bound states lie between
$\,(m_1\!+\!m_2\!+\!m_3\!-2\, \hbox{Inf}(m_i))\,$ (i.e. above the highest
negative-energy threshold) and $\,(m_1\!+\!m_2\!+\!m_3),\,$ there is no
degenerescence with the "one plus-two minus" states. On the contrary, no
weak field assumption could prevent the degenerescence with the "two
plus-one minus" states.\par
 The continuum dissolution problem is avoided with the potentials containing
the positive-energy projectors 
$\,\Lambda^{++}_{ij},\,$ or their covariant substitutes
$\,\theta^{++}_{ij}.\,$ The 3D equation built with Salpeter's potentials
$\,V^{S}_{ij},\,$ which include a $\,\tau_{ij}\,$ at left and a
$\,\tau^2_{ij}\,$ at right, can be splitted into two
uncoupled equations in the $\,\tau_1\tau_2\tau_3\!=\!\pm1\,$ subspaces
and the solutions are eigenstates of $\,\tau_1\tau_2\tau_3\,$ and
free of continuum dissolution. The $\,\tau_1\tau_2\tau_3\!=\!+1\,$
subspace includes the three-plus and the two-minus one-plus components. The
interaction term is hermitian for a scalar product built with
$\,\tau_{123}=\hbox{sign\,}(\tau_1+\tau_2+\tau_3).\,$ Equation
(\ref{42**})  becomes then a logical three-fermion extension of Salpeter's
equation and exhibits a particle-antiparticle symmetry. We do not expect
however to get a large contribution from the two-minus one-plus components.

\section{Kernel-based 3D reductions.}
\subsection{Two fermions.}
Our 3D reduction of the two-fermion Bethe-Salpeter equation was built around
an approximation of the free propagator. However, the first
step of our 3D reduction of the three-fermion Bethe-Salpeter equation is
built around instantaneous approximations of the kernels, while the second
step is again built around approximations of the free propagators. It is
therefore interesting to examine also how a 3D reduction of the two-fermion
Bethe-Salpeter equation can be built around an instantaneous approximation of
the kernel. Let us thus consider an approximation $\,K^Z\,$ of
the Bethe-Salpeter kernel:
\begin{equation}K\,=\,K^Z\,+\,K^R.\label{*28}\end{equation} The Bethe-Salpeter
equation becomes
\begin{equation}\Phi=G^0K^Z\Phi+G^0K^R\Phi\label{*29}\end{equation}  
\begin{equation}\Phi=(1-G^0K^R)^{-1}G^0K^Z\Phi=G^KK^Z\Phi\label{*30}
\end{equation}
with
\begin{equation}G^K=\,G^0\,+\,G^0K^RG^0+G^0K^RG^0K^RG^0+\cdots\equiv\,
G^0+G^{KR}\,.\label{*31}\end{equation} If we now specialize
$\,K^Z\,$ to an instantaneous positive-energy kernel ($K^Z\,$ independent of
$\,p_0\,$ and equal to
$\,\beta_1\beta_2\Lambda^{++}\beta_1\beta_2K^Z\Lambda^{++}\,$), eq. (\ref{*30}) 
leads to the 3D equation
\begin{equation}\phi\,=\,(g^0+g^{KR}\,)\,V^Z\,\phi\label{*32}\end{equation} with
\begin{equation}\phi=\Lambda^{++}\int dp_0\,\Phi(p_0),\qquad
V^Z=-2i\pi\,\beta_1\beta_2K^Z,
\label{*33}\end{equation}
\begin{equation}g^{KR}=-{1\over2i\pi}\,\Lambda^{++}\,\int
dp'_0\,dp_0\,G^{KR}(p'_0,p_0)\,\beta_1\beta_2\,\Lambda^{++}.\label{*34}
\end{equation}
The first-order energy shift is, with the replacement
$\,g^{KR}V^Z\phi\approx g^{KR}(P_0\!-\!S)\phi:$  
\begin{equation}P_0-P^{(0)}_0=-{1\over2i\pi}\,<\,(P^0\!-\!S)\int
dp'_0\,dp_0\,G^0(p'_0)K^R(p'_0,p_0)G^0(p_0)\,\beta_1\beta_2\,(P^0\!-\!S)\,>.
\label{*35}\end{equation} Here again, all elements of (\ref{*35}) are
total-energy dependent. Although (\ref{*35}) is symmetric, the perturbation
potential
$\,(P_0\!-\!S)g^{KR}V^Z\,$ of (\ref{*32}) is not. This does not exclude the
possibility of a real energy spectrum: if we admit an energy dependence of the
potential (which we cannot avoid), we have also the possibility of performing
an infinity of rearrangements of the equation. We can for example render the
potential symmetric by treating $\,g^{KR}\,$ as a perturbation of
$\,g^0\,$   (like $\,G^R\,$ as a perturbation of $\,G^{\delta}\,$ in  
section (2.2)). We can equivalently start with the corresponding 4D equation
(\ref{*30}):
\begin{equation}\Phi=(G^0+G^{KR})\,K^Z\Phi=\phi+G^{KR}\,K^Z\Phi\label{*36}
\end{equation}
\begin{equation}\phi=G^0K^Z\Phi\,=G^0K^Z(1-G^{KR}\,K^Z)^{-1}\,\phi=G^0K^K
\phi\label{*37}
\end{equation} where
\begin{equation}K^K\,=K^Z\,+\,K^ZG^{KR}K^Z\,+\,\cdots\label{*38}\end{equation}
 is a symmetric positive-energy instantaneous potential, since it begins and
ends with
$\,K^Z\,$. The 3D equation is then
\begin{equation}\eta\,=\,g^0\,V^K\,\eta,\qquad
V^K\,=\,-2i\pi\,\beta_1\beta_2\,K^K,\qquad
\eta=\int dp_0\,\phi\,(p_0)\label{*39}\end{equation} and the first-order energy
shift is still given by (\ref{*35}). As for the propagator-based
reduction, we can represent the various contributions by Feynman graphs. We can
indeed get
$\,K^K\,$  by considering the transition operator
\begin{equation}T\,=\,(K^Z+K^R\,)\,+\,(K^Z+K^R\,)\,G^0\,(K^Z+K^R\,)\,+\,\cdots
\label{*40}\end{equation} and skipping some terms according to the two
rules:\par -- Each term must begin and end with
$\,K^Z$\par -- There must be at least one $\,K^R$ between two consecutive
$\,K^Z.$
\par\noindent  The 3D transition operator is now
$\,-2i\pi\beta_1\beta_2T^K\,$ with 
$$T^K=K^K+K^KG^0K^K+\cdots=K^K(1-G^0K^K)^{-1}$$
$$=K^Z(1-G^{KR}K^Z)^{-1}(1-G^0K^Z(1-G^{KR}K^Z)^{-1})^{-1}$$
$$=K^Z(1-G^{KR}K^Z-G^0K^Z)^{-1}=K^Z(1-G^KK^Z)^{-1}$$
$$=K^Z(1-(1-G^0K^R\,)^{-1}G^0K^Z\,)^{-1}$$
$$=K^Z(1-G^0K^R-G^0K^Z\,)^{-1}(1-G^0K^R\,)=K^Z(1-G^0K\,)^{-1}(1-G^0K^R\,)$$
\begin{equation}K^Z(1+G^0T\,)\,(1-G^0K^R\,)\label{*41}\end{equation} Replacing
then
$\,K^R\,$ by $\,K\!-\!K^Z\,$ and
$\,T(1\!-\!G^0K)\,$ by $\,K\,$ gives
\begin{equation}T^K\,=\,K^Z+K^ZG^0K^Z+K^ZG^0\,T\,G^0K^Z.\label{*42}\end{equation}
In contrast with what happens in the propagator-based reductions, the on
mass shell restriction of this operator is not the physical scattering
amplitude of field theory. In fact, the Bethe-Salpeter equation (\ref{1}),
without inhomogeneous term, is a bound state equation, and for the bound states
we are only interested in the poles of (\ref{*42}). These poles are the poles of
$\,T,\,$ unless the integrations on
$\,p'_0,p_0\,$ make some residues vanish. Getting also the correct physical
scattering amplitudes was a welcome bonus of the propagator-based
reductions.\par  If we want
$\,T\,$ in terms of
$\,T^K,\,$ we can write (\ref{*41}) as
$$T^K(1-G^0K^R\,)^{-1}\,=\,K^Z(1+G^0T\,)\,=\,(K-K^R\,)(1+G^0T\,)\,
$$
\begin{equation}=\,T-K^R\,(1+G^0T\,)\,=\,(1-K^R\,G^0\,)T\,-K^R\label{*42b}
\end{equation} 
$$T\,=\,(1-K^R\,G^0\,)^{-1}\,T^K\,(1-G^0K^R\,)^{-1}\,+\,(1-K^R\,G^0\,)^{-1}
\,K^R$$
\begin{equation}=\,(1+T^RG^0)\,T^K\,(1+G^0T^R)\,+\,T^R\label{*43}\end{equation}
with
\begin{equation}T^R\,=\,K^R+K^RG^0K^R+\cdots.\label{*44}\end{equation} We have a
large freedom in the choice of $\,K^Z.\,$ We could choose
\begin{equation}K^Z\,=\,\beta_1\beta_2\,\Lambda^{++}\,\beta_1\beta_2\,K(s,s)\,
\Lambda^{++}.\label{*45}\end{equation} We could also keep only the ladder term
of $\,K,\,$ or a part of it in (\ref{*45}), like the Coulomb part of the
one-photon exchange contribution. We could use $\,K^T\,$ (in this case we
expect that the remainder of
$\,K^K\,$  does not finally contribute to the total energy spectrum) or an
approximation of it (one and two photon exchange terms, for example). Phillips
and Wallace
\cite{21} suggested to choose $\,K^Z\,$ in order to annihilate
$\,g^{KR}\,$ exactly or up to a given order. If we choose to annihilate it at
first-order, the first-order energy shift (\ref{*35}) vanishes and we get
\begin{equation}K^Z={-1\over(2\pi)^2}\,\beta_1\beta_2\,(g^0)^{-1}\Lambda^{++}
\int
dp'_0\,dp_0\,G^0(p'_0)K(p'_0,p_0)G^0(p_0)\,\beta_1\beta_2\,(g^0)^{-1}
\Lambda^{++}.
\label{*46}\end{equation} The skipping of terms in (\ref{*40}) is then improved,
as we must now have at last {\it two}
$\,K^R$ between two consecutive
$\,K^Z.\,$ If we choose to annihilate $\,g^{KR}\,$ exactly, we get, using
(\ref{*43}), (\ref{*34}) and the relation
$\,G^{KR}\!=G^0T^RG^0:\,$
\begin{equation}{-1\over(2\pi)^2}\,\beta_1\beta_2\,(g^0)^{-1}\Lambda^{++}\int
dp'_0\,dp_0\,G^0(p'_0)T(p'_0,p_0)G^0(p_0)\,\beta_1\beta_2\,(g^0)^{-1}
\Lambda^{++}
=T^K\label{*46c}\end{equation} which is the $\,\Lambda^{++}\,$ projection of the
expression given in
\cite{21} (they use Salpeter's propagator and write thus
$\,\Lambda^{++}\!-\!\Lambda^{--}\,$ instead of
$\,\Lambda^{++}\,$). In this case, $\,T^K\,$ gives the correct physical
scattering amplitudes when $\,P^0\!\to\! E_1\!+\!E_2,\,$ as the operators
$\,(g^0)^{-1}\,$ kill the contributions of the singularities other than the
poles of $\,G^0.$

\subsection{Three fermions.}
We could also use the results of the kernel-based 3D reduction of
the two-fermion Bethe-Salpeter equation as the starting point of a
3D reduction of the three-fermion Bethe-Salpeter equation, by
choosing
\begin{equation}K^0_{ij}\,=\,K^K_{ij}(s,s,P_0\!-\!h_k)\label{127*}\end{equation}
as starting approximations in (\ref{*58}), in order to recover, at
the 2+1 separated clusters limits, the kernel-based 3D equations.           
 
\subsection{One-photon exchange.}    Let us now compute (\ref{*46}) with $\,K\,$
given by the one-photon exchange graph, in Feynman gauge:
\begin{equation}K\,=\,{2ie^2\over(2\pi)^3}\,\,\,{\gamma_1\!\cdot\gamma_2
\over(p'\!-\!p)^2
+i\epsilon}.\label{*47}\end{equation}
We have
\begin{equation}K^Z\,=\,-\,{2ie^2\over
(2\pi)^5}\,\beta_1\beta_2\,\Lambda^{++}\,\gamma_1\!\cdot\gamma_2\,\Lambda^{++}\,
I\label{*47bis}\end{equation} 
 With
$$I=\int dp'_{10}\,dp_{10}\quad{P_0-E'_1-E'_2\over (p'_{10}-E'_1+i\epsilon'_1)\,
(P_0-p'_{10}-E'_2+i\epsilon'_2)}\,\,$$
\begin{equation}{1\over (p'_{10}-p_{10})^2-\vec k^2+i\eta}\quad{P_0-E_1-E_2\over
(p_{10}-E_1+i\epsilon_1)\,(
P_0-p_{10}-E_2+i\epsilon_2)}\label{*48}\end{equation} 
and $\,k\!=\!p'\!-\!p.\,$   
Closing the integration paths clockwise, and taking the $\,\epsilon_1\to 0\,$
limit before the $\,\eta\to 0\,$ limit in all terms, we get
\begin{equation}I=-4\pi^2\,\, {1\over (E'_1-E_1)^2-\vec
k^2}\,\,\,\,(\,1\,+\,{1\over2\vert\vec
k\vert}\,R_K\,),\label{*48bis}\end{equation}    
\begin{equation}R_K\,=\,{(E'_1-E_1-\vert\vec
k\vert\,)(P_0-E_1-E_2)\over P_0-E'_1-E_2-\vert\vec
k\vert\,}\,-\,{(E'_1-E_1+\vert\vec k\vert\,)(P_0-E'_1-E'_2)\over
P_0-E_1-E'_2-\vert\vec k\vert\,}.\label{*50}\end{equation} The first
term comes from the residues of the poles of the propagators only, while the 
terms of $\,R_K\,$ combine the residues of one pole of the propagators and one
pole of the kernel. Modifying the integration paths or the order of the limits
would give the same result (after rearrangements). At lowest order, we get
\begin{equation}R_K\,\approx\,2P_0-E'_1-E'_2-E_1-E_2.\label{*51}\end{equation}
The contribution of $\,R_K\,$ to the bound state energy will finally be in
$\,\alpha^3,\,$ while the replacement of $\,(E'_1-E_1)^2-\vec k^2\,$ by simply 
$\,-\vec k^2\,$ (Coulomb potential) would lead to a correction in
$\,\alpha^4.\,$ This means that the relative energy dependence of the kernel, at
the ladder approximation, contributes mostly by providing supplementary poles
inside the integration paths. Beyond the ladder approximation however, the
contribution of the first crossed graph will cancel the leading term in
$\,\alpha^3,\,$ leaving a contribution in $\,\alpha^4\,$ \cite{38*}. This is
not a surprise, as it is  well known  that, in well chosen propagator-based
reductions, all higher-order contributions cancel mutually at the one-body limit
\cite{3,7}.\par
By rearranging (\ref{*48bis}, \ref{*50}), we can also write
\begin{equation}I\,=\,{-4\pi^2\over2\vert\vec k\vert}
\,\left( {1\over P_0-E'_1-E_2-\vert\vec
k\vert}\,+\,{1\over P_0-E_1-E'_2-\vert\vec
k\vert}\right)\label{*51b}\end{equation}
which is the expression obtained in time-ordered perturbation theory \cite{51}.
 Here, each term describes the emission of a photon by one fermion,
followed by its absorption by the other one.

\subsection{Time-ordered perturbation theory}
In time-ordered perturbation theory (TOPT), one takes the Fourier
transform of the full propagator (as given by the Feynman graphs
method) with respect to the energy variables. Each graph becomes
then splitted into several time-ordered graphs, and one integrates
the result with respect to the time intervals. This leads to TOPT
"Feynman rules". The TOPT graphs can be arranged in order to get
an iterative equation for the propagator, and, by isolating the
residues of the three-body bound state poles, to get an
homogeneous 3D equation for the three-body bound state wave
functions (this imitates the usual derivation of the inhomogeneous and
homogeneous Bethe-Salpeter equations). The cluster-separability
property is translated by the factorization of the propagator at
the different vanishing interactions limits. The two and three-body 3D
potentials are directly given, at any order, by the TOPT
"Feynman rules". Up to now, this approach has been applied to a
very simple model: the exchange of a scalar meson between
two or three positive-energy bosons or fermions at the ladder
approximation \cite{51}. Our results in this work can be applied to more
general interactions but should lead to more complicated calculations than
TOPT at higher-orders. A comparison of our first-order contributions computed
with the simple Bethe-Salpeter kernel used in \cite{51} should be
interesting.\par It must be noted that our motivation for introducing the
positive-energy projectors is not that of \cite{51}: we make the choice of
approaching the exact equation for the positive-energy components, while
\cite{51} simply neglects these components as a temporary simplification. It
will be interesting to see in which form the continuum dissolution problem 
appears in TOPT when the negative-energy components  are taken into
account.

\section{Relation with Faddeev formalism and Gross' spectator model.}
The nonrelativistic Faddeev equations can be obtained by transforming
Schr\"odinger's equation (in a first step, without three-body terms) into
a set of three coupled equations for three parts $\,T_{ij}$ of the
transition operator ($\,T_{ij}$ denotes the contribution of all graphs
beginning by a (ij) interaction). The input is the set of the three
two-body transition operators and the resulting series expansion contains
only connected graphs (never twice the same two-body transition
operator). This formalism is well adapted to the description of the
various scattering processes, such as
$\,(12)+3\to (12)+3
$ (elastic scattering),
$\,(12)+3\to (12)^*+3 $
 (excitation), $\,(12)+3\to 1+(23) $ (rearrangement), $\,(12)+3\to 1+2+3
$ (breakup). The (123) bound states correspond to the poles of this
transition operator. \par The structure of these equations can be
generalized to relativistic equations which are not necessarily (exactly)
reducible to a single 3D equation. \par In Gross'  spectator model
\cite{12,13}, Faddeev's type equations are deduced from the
Bethe-Salpeter equation and the transition operator is written in terms
of the two-body transition operators. The relative time variables are
then eliminated by putting, in each three-body  propagator, all the
"offmassshallness" on the only fermion which interacts before and after. 
The Lorentz invariance-cluster separability requirement is satisfied by
applying suitable Lorentz boosts on the two-body transition operators.
\par The main difference between our approach and Gross' approach comes
from the fact that we are (presently) interested by the (123) bound
states only, to be computed principally by using a single 3D equation
(such as the one from which Faddeev starts). Instead of working with the
two-body transition operators, we work with the two-body potentials. We
can however present our approach in terms of the two-body transition
operators. Our approximation of each two-body transition operator is then
unique and does not depend on the operators which come in front and
behind it in the expansion of the Faddeev equations. This enables us to
describe our model by a single 3D potential equation, by a set of three
Faddeev equations, or by the expansion of the three-body transition
operator in terms of the two-body ones. Another feature of our model is
to not introduce Lorentz boosts by hand: the Lorentz-invariance / cluster
separability requirement is exactly satisfied if we do not truncate our
potentials (practically, this means that the satisfaction of this
requirement and the approximation of the potentials can be improved
together).  Let us go back to the three-fermion Bethe-Salpeter equation 
and neglect for simplicity the three-body irreducible kernel $\,K_{123}.
\,$ Defining
\begin{equation}\Phi_{12}\,=\,G^0_1G^0_2\,K_{12}\,\Phi,\cdots\qquad\,
\Phi\,=\,\Phi_{12}\,+\,\Phi_{23}\,+\,\Phi_{31}\,, 
\label{64}\end{equation} we get
\begin{equation}(1\,-\,G^0_1G^0_2\,K_{12})\,\Phi\,=\,\Phi_{23}\,+\,
\Phi_{31} 
\label{65}\end{equation}
\begin{equation}\Phi\,=\,(1\,+\,G^0_1G^0_2\,T_{12})\,(\,\Phi_{23}\,+\,
\Phi_{31}\,) 
\label{66}\end{equation}
\begin{equation}\Phi_{12}\,=\,G^0_1G^0_2\,T_{12}\,(\,\Phi_{23}\,+\,
\Phi_{31}\,). 
\label{67}\end{equation} Writing then
\begin{equation}\Phi_{12}\,=\,G^0_1G^0_2G^0_3\,\beta_1\beta_2\beta_3\,
\phi_{12},\,\cdots\label{68}\end{equation}
in order to factor out the propagators, we get
\begin{equation}\phi_{12}\,=\,\beta_1\beta_2\,T_{12}\,G^0_1G^0_2\,\beta_1
\beta_2\,(\,\phi_{23}\,+\,\phi_{31}\,).
\label{69}\end{equation}\hfill\break   The (12) transition matrix element
corresponding to our approximation (\ref{*58}) of $\,K_{12}\,$ is
$$T_{12}(\,p'_{120},\,p_{120},\,P_0-p_{30}\,)\,\approx\,T_{12}^0(p_{30})
\,$$
\begin{equation}\,\,\equiv\,\,\,K^{T++}_{12}(s_{12},s_{12},P_0\!-\!h_3)
(1-G^0_1G^0_2K^{T++}_{12}(s_{12},s_{12},P_0\!-\!h_3))^{-1}.\label{70}
\end{equation}
This $\,T_{12}^0(p_{30})\,$ is  analytical in the Im$(p_{30})<0\,$ half
plane and 
\begin{equation}T_{12}^0(h_3)\,=\,T^{++}_{12}(s_{12},s_{12},P_0\!-\!h_3).
\,\label{71}\end{equation}   
This approximation, combined with equation (\ref{69}), implies that
$\,\phi_{12}(p_{120},p_{30})\,$ is independent of
$\,p_{120}\,$ (let us write thus $\,\phi_{12}(p_{30})\,$). Equation
(\ref{69}) becomes then
\begin{equation}\phi_{12}(p_{30})\,=\,\beta_1\beta_2\,T_{12}^0(p_{30})\,
\int
d\,p_{120}\,
G^0_1G^0_2\,\beta_1\beta_2\,[\,\,\phi_{23}(p_{10})\,+\,\phi_{31}(p_{20})
\,]\label{72}\end{equation}                 
where $\,p_{10},p_{20}\,$ must be written in terms of
$\,P_0,p_{30},p_{120}.\,$ Equation (\ref{72}), together with similar
equations for $\,\phi_{23}(p_{10})\,$ and $\,\phi_{31}(p_{20}),\,$ admits
solutions which are analytical in the Im$(p_{k0})<0\,$ half planes. At
$\,p_{k0}=h_k,\,$ we get the 3D Faddeev equations
\begin{equation}\phi_{12}(h_3)\,=\,T^{\,3D++}_{12}(P_0-h_3)\,{1\over
P_0-S+i\epsilon P_0}\,[\,\,\phi_{23}(h_1)\,+\,\phi_{31}(h_2)\,]\,,\cdots
\label{73}\end{equation} which can easily be transformed back into our
basic three-cluster potential equation (by performing in the reverse
order the transformations made above at the 4D level). Note that we would
get the same result by approaching directly              
$\,T_{12}\,$ by $\,T_{12}^0(h_3)\,$ instead of $\,T_{12}^0(p_{30}).\,$
This means approaching $\, K_{12}\,$ by a given function
$\,K_{12}^0(p_{30}),\,$ analytical in the Im$(p_{30})<0\,$ half plane and
equal to $\,K^{T++}_{12}(s_{12},s_{12},P_0\!-\!h_3)\,$ at $\,p_{30}=h_3.$
\par The key point of the manipulations above is the dominance of the
positive-energy poles of $\,G^0_1G^0_2\,$ in (\ref{69}). This was
obtained by approaching $\,K^T_{12}\,$ (or $\,T_{12}\,$) by a constant
with $\,\Lambda^{++}_{12}\,$  positive-energy projectors. We could try to
insure the dominance of these two poles more economically. Let us come
back to equation (\ref{69}) without making any approximation on
$\,T_{12}.\,$ The elements of (\ref{69}) are then the operator and
functions
\begin{equation}T_{12}\,(\,p'_{120},\,p_{120},\,
P_0-p_{30})\label{74}\end{equation} 
\begin{equation}\phi_{12}\,(\,p'_{120},\,p_{30}\, ),\quad
\phi_{23}\,(\,p_{230},\,p_{10}\, ),\quad\phi_{31}\,(\,p_{310},\,p_{20}\,
)\label{75}\end{equation} and (\ref{69}) must be integrated with respect
to $\,p_{120},\,$ with $\,p_{30}\,$ fixed. We must thus write
$\,p_{230},\,p_{10},\,p_{310},\,p_{20}\,$ in terms of
$\,p_{120},\,p_{30}.\,$ Searching as above for solutions
$\,\phi_{ij}\,(\,p'_{ij0},\,p_{k0}\, )\,$ which are analytical in the
Im$(p_{k0})<0\,$ half planes, we shall close our integration path
clockwise (counterclockwise) in front of $\,\phi_{23}\,$
($\,\phi_{31}\,$) and keep only the residue of the pole of
$\,\Lambda^{++}_{12}G^0_1G^0_2\,$ which puts fermion 1 (2) on its
positive-energy mass shell. The elements of (\ref{69}) are then replaced by
\begin{equation}T_{12}\,(\,p'_{120},\,p_{120},\, P_0-p_{30})\,\to\,\,
T_{12}\,(\,p'_{120},\,s_{12}\,\mp\,{P_0-S\over2}\,\pm\,{p_{30}-h_3\over2},
\,
P_0-p_{30})\,\label{76}\end{equation}                     
\begin{equation}\int
d\,p_{120}\,G^0_1G^0_2\,\beta_1\beta_2\,\to\,{-2i\pi\,\Lambda^{++}_{12}\over
(P_0-S)-(p_{30}-h_3)+i\epsilon}\label{77}\end{equation}
\begin{equation}\phi_{23}\,(\,p_{230},\,p_{10}\,
)\,\to\,\phi_{23}\,(\,s_{23}\,+\,{P_0-S\over2}\,-\,(\,p_{30}-h_3\,),
\,E_1\,)\label{78}\end{equation}  
\begin{equation}\phi_{31}\,(\,p_{310},\,p_{20}\,
)\,\to\,\phi_{31}\,(\,s_{31}\,-\,{P_0-S\over2}\,+\,(\,p_{30}-h_3\,),
\,E_2\,)\label{79}\end{equation}      
where we take the upper signs  in (\ref{76}) in front of $\,\phi_{23},\,$
the lower signs in front of
$\,\phi_{31}.$\par The manipulations above are submitted to some
restrictions on the $\,T_{ij}.\,$  In the $\,p_{k0}\,$ variable the
$\,T_{ij}\,$ must be analytical in the Im$(p_{k0})<0\,$ half planes. In
the $\,p'_{ij0}\,$ and
$\,p_{ij0}\,$ variables the $\,T_{ij}\,$ must be analytical in the whole
complex plane. Moreover, the 
$\,T_{ij}\,$ must also be asymptotically bounded in the three variables 
and contain $\,\Lambda^{++}_{ij}\,$ positive-energy projectors. These
conditions are of course not satisfied by the exact transition matrix
elements, but we shall assume that the singularities and the
negative-energy parts of the
$\,T_{ij}\,$ can be neglected in the computation of the integrals with
respect to the relative energies. \par            Taking the equation for
$\,\phi_{12}\,$ at $\,p_{30}\!=\!E_3\,$ and
$\,p'_{120}=s^{\pm}_{12}=s_{12}\,\pm\,{1\over2}(P_0-S),\,$ means
that we put fermions 2 and 3 (for +) or fermions 1 and 3 (for -) on their
mass shells. This leads to 
\begin{equation}\phi^2_{12}\,=\,-2i\pi\,\beta_1\beta_2\,(\,T^{\,2,1}_{12}
\,{1\over
P_0-S+i\epsilon}\,\,\phi^2_{23}\,+\,T^{\,2,2}_{12}\,{1\over
P_0-S+i\epsilon}\,\,\phi^1_{31}\,)\label{80}\end{equation}
\begin{equation}\phi^1_{12}\,=\,-2i\pi\,\beta_1\beta_2\,(\,T^{\,1,1}_{12}
\,{1\over
P_0-S+i\epsilon}\,\,\phi^2_{23}\,+\,T^{\,1,2}_{12}\,{1\over
P_0-S+i\epsilon}\,\,\phi^1_{31}\,)\label{81}\end{equation} where the upper
indexes refer to the initial and final on mass shell non-spectator fermions.
Equations (\ref{80}),(\ref{81}), together with similar equations for the
(23) and (31) pairs, are a closed  system of six 3D equations. In each
$\,T_{ij}\,$ two fermions are on the mass shell: the spectator fermion k
and the fermion which is not going to interact next left ($\,p'_{ij0}\,$)
or next right ($\,p_{ij0}\,$). This is of course the philosophy of Gross'
spectator model \cite{12,13}. \par If we replace $\,s^{\pm}\,$ by $\,s\,$
in the $\,T_{ij}\,$ we get our basic three-cluster model. This
supplementary approximation can be justified by noticing that we already
neglected the contributions of the singularities of
$\,T_{ij}\,$ in the $\,p'_{ij0}\,$ and $\,p_{ij0}\,$ complex planes. The
dependence on these variables can thus not be very strong, as we know
that a function which is analytical and bounded on the whole complex
sphere must necessarily be a constant. Using the same argument of the
consistency of the approximations, we could also argue that $\,T_{ij}\,$
constant implies $\,K^T_{ij}\,$ and $\,K_{ij}\,$ constant with
$\,K^T_{ij}=K_{ij}.\,$ Our basic three-cluster equation is thus not a
priori a worse approximation of the three-fermion Bethe-Salpeter equation
than Gross' equations, nor a better approximation than the simpler
positive-energy instantaneous approximation or the corresponding Born
approximation. It reflects the choice of preserving the exact
cluster-separated limits. \par 
To conclude the comparison between our 3D reduction and Gross' spectator
model, we can say that, at lowest order, our 3D reduction is much simpler:
our two-body transition operators are always taken at the same value of the
corresponding relative energie, so that we can transform the set of coupled
equations for the transition operators into a unique equation for a wave
function. We do not use Lorentz boosts, as we choose to exploit the implicit
Lorentz covariance of the unapproximated separated-clusters limit equations,
so that the effects of these boosts will be approached in principle by the
progressive introduction of the higher-order two-fermion terms.  Gross'
higher-order corrections introduced by taking the transition operators with
different fermions on the mass shell are not a priori more important than the
contributions of the singularities of these transition operators themselves to
the closed path integrals in the relative energies. Our brief calculation with
the one-photon exchange graph (subsection 5.3) shows that, in this case, the
contributions of these singularities are in fact more important, by an order
of $\,\alpha\,$ (but it turns out that the leading order of these
contributions will be cancelled by the leading order contribution of the first
crossed graph
\cite{38*}). Our higher-order calculation takes both types of
contributions into account.

\section{Heavy mass limits and external potentials.}
\subsection{Heavy mass limits in the two-fermion problem.} When the mass
of one of the fermions goes to infinity, we must find the equation of the
other fermion in a potential (a Dirac equation with a Coulomb potential
in QED). This limit can be obtained directly, or via a rearrangement of
the equation. With Breit's equation with the second-order constraint and
with Sazdjian's equation, the higher-order ladder and crossed terms
cancel mutually at the one-body limit, so that the correct limit of the
potential is already contained in the Born term. With Salpeter's equation
with the second-order constraint, and with the positive-energy equation with a
second-order constraint, the limit is also a Dirac-Coulomb equation, but solved
with respect to the
$\,\Lambda^+\psi\,$ part of the wave function: the
$\,\Lambda^-\psi\,$ part is transformed into higher-order contributions
to the potential. The physical content remains identical to that of a
Dirac-Coulomb equation, but the correct limit of the potential is no more
entierely contained in the Born term, so that a truncation of the
potential would spoil the one-body limit [8-12].
With the first-order constraint, the results are similar, but after
complicated rearrangements of the equation. \par
In the three-fermion problem,
we expect similarly that, at the limit of an infinite fermion mass, we would
get an equation describing two fermions in an external potential. It is thus
interesting to compare this equation with the equation we wrote
directly \cite{11}  for the two-fermion in an external potential system.

\subsection{Two fermions in an external potential.}

 The two-fermion plus potential problem can be approached in two ways
which we shall call the two-body and the three-body approaches. In the
two-body approach, the external potential is included in the definition
of the creation and annihilation operators of field theory. Practically,
we can keep the equations obtained in the treatment of the pure
two-fermion problem,  using simply a generalized definition of the
$\,h_i$  \cite{9,10,11}:
\begin{equation}h_i = \vec \alpha_i\, . \vec p_i + \beta_i\,
m_i\,+\,V_i(\vec x_i)\label{51}\end{equation} where $\,V_i\,$ is the
external potential acting on the fermion i. 
 All quantities can be expanded on the basis built with the eigenstates of
$\,h_1\,$ and $\,h_2.$ In the three-body approach, the free creation and
annihilation operators are used and the external potential is treated as
an heavy third body. This approach will be presented as an heavy mass
limit of the three fermion problem in the next subsection.\par 
The equations for two fermions in an external potential must be written
in the laboratory reference frame (in which the external potential is
defined) and are not Lorentz invariant. If we "switch off" the mutual
interaction we get a pair of uncoupled equations for two independent
fermions in an external potential. If we switch off the external
potential the equations remain written in the laboratory frame, which
still refers to the vanished external potential. If the reduction series
is not truncated, this no-external interaction limit of the equation is
equivalent to a pure two-fermion covariant Bethe-Salpeter equation. 

\subsection{Heavy mass limits in the three-fermion problem.}
Let us now take our 3D equation (\ref{42**}), with the potentials
$\,V_{ij}\,$ obtained by performing 3D reductions based on the
$\,G_{ij}^{\delta P}\,$ as in section 2, but with the second-order constraints
$\,\delta(p_{ij0}\!-\!\mu_{ij})\,$ instead of the first-order constraints
$\,\delta(p_{ij0}\!-\!s_{ij}).\,$     
 When $\,m_3,$ for example, goes to infinity (two-body limit), we get,
writing $\,P_0=W_{12}+m_3:\,$ 
\begin{equation}\,\psi\,=\,{1\over W_{12}-h_1-h_2}\,(V_{12}+V^+_2
+V^+_1)\,\psi\label{52}\end{equation} with
$\,(P_0-h_3)\,$ replaced by $\,W_{12}\,$ in $\,V_{12}.\,$ The
potential
$\,V^+_2\,$ is given by the series
\begin{equation}V^+_2\,=\,\Lambda^+_2\left[V_2\,+\,V_2\,{\Lambda^-_2\over
(W_{12}\!-\!h_1)+E_2-i\epsilon}\,V_2\,+\,...\right]\Lambda^+_2\label{53}
\end{equation}
where
$\,V_2\,$ is an external potential acting on fermion 2 (it is a Coulomb
potential in QED, if we use the second-order constraint,  but in general it
could still depend on $\,(W_{12}-h_1)\,$). The potential
$\,V^+_1\,$ is given by a similar formula.\par\noindent The differences
with the two-body approach of the two-body plus potential problem
are:\par\hfill\break
\indent -- The projectors $\,\Lambda^{\pm}_i\,$ are the free ones.\par --
The external potential for fermion 2 is now $\,V^+_2,\,$ given by the series
(\ref{53}). This is the potential obtained by solving the equation for
fermion 2 in the external potential $\,V_2\,$ with respect to the positive
free energy components, the unobservable energy of fermion 2 being replaced
in it by $\,W_{12}\!-\!h_1.$ 
\par --$\,V_2\,$ itself could still depend on
$\,(W_{12}\!-\!h_1)\,$ and vice-versa. \par\hfill\break\noindent The cluster
separability property in the three-body way survives the high-mass limit:
switching off $\,V_2\,$ and $\,V_1\,$ leads to the equation for the (12)
two-fermion system, switching off $\,V_{12}\,$ and $\,V_1\,$ leads
to a free Dirac equation for fermion 1 with the $\,\Lambda^+_2\,$
projection of the Dirac equation for fermion 2 in the external potential
$\,V_2.\,$ In the two-body approach of the two-fermion plus potential
problem we had however more: switching off the mutual interaction only
led to a pair of independent equations for each fermion in the external
potential. Here, in the three-body approach, the equation does not split
perfectly into two parts: we can write $\,W_{12}\!=\!W_1+W_2,\,$ but we
have 
$\,W_{12}-h_1\,$ instead of $\,W_2\,$ in the series defining $\,V^+_2\,$
and vice-versa. This is a consequence of the energy dependence introduced
into (\ref{53}) by the use of the anti-continuum dissolution projectors,
combined with our choice of putting the spectator fermion on the mass
shell in each two-body interaction, and neglecting the three-body terms
which could balance this modification. The discrepancy is of order
$\,V^4.\,$ At the two-fermion plus potential level, a better but still
not perfect equation would be obtained by replacing $\,h_1\,$ by
$\,h_1+V_1\,$ in the series defining $\,V^+_2\,$ and vice-versa.

\section{Conclusions}
 We have written a 3D equation for three fermions (our basic
three-cluster equation)  by combining the three two-body potentials
obtained by an untruncated propagator-based 3D reduction of the corresponding
two-fermion Bethe-Salpeter equations, putting the spectator fermion on the
mass shell. In this way, the cluster-separated limits are exact, and
the   Lorentz invariance / cluster separability requirement is
automatically satisfied, provided no supplementary approximation, like
the Born approximation, is made. Such a truncation would render the
Lorentz covariance of the two-fermion clusters only approximate (a Born
approximation preserving the  Lorentz invariance / cluster separability
property can be obtained by using  another 3D reduction based on a covariant
second-order two-body propagator of Sazdjian, combined with a covariant
substitute of
$\,\Lambda^{++}.$ This leads to a 3D three-cluster equation which is 
covariantly Born approximable, but more complicated
\cite{20}\,). Our 3D equation
can be written  in terms of the two-body potentials, or in terms of two-body
transition operators (Faddeev formalism). The use of positive free-energy
projectors in the chosen reductions of the two-fermion Bethe-Salpeter
equations   prevents our 3D three-fermion equation from continuum
dissolution. The potentials are hermitian and depend only slowly on the total
three-fermion energy.  The one high-mass limits of our "basic three-cluster
equation" are approximately exact. The correction of the remaining
discrepancy would demand the introduction of higher-order three-body
terms.    
\par  Our combination of cluster separability and Lorentz invariance in
the three fermion problem makes explicit use of the fact that the
clusters can only be two-fermion and/or free fermion states. This is not
directly adaptable to four or more fermion systems. In this respect, 3
is still not N.\par   Our 3D equation can be written directly, or derived
from an approximation of the three-fermion Bethe-Salpeter equation, in
which the three-body kernel is neglected while the two-body kernels are
approached by positive-energy instantaneous expressions, with the
spectator fermion on the mass shell. The correction terms are thus known
at the Bethe-Salpeter level and can be transformed into corrections to
the 3D equation.\par
Our lowest-order three-fermion 3D equation is much simpler than the
equations written in the framework of Gross' spectator model, which
contain some induced higher-order three-fermion contributions. Our
calculation of the next order shows that these contributions are not a
priori more important that the ones coming from the singularities of the
kernels, neglected in Gross' model. We think that the relative
importance of the different contributions could largely depend on the
specific problem studied.\par 
  There exists an infinity of ways of performing the 3D reduction of the
two-fermion Bethe-Salpeter equation. The potentials generated by these
reductions could all by used to build a three-fermion 3D equation,
keeping however in mind the continuum dissolution problem. This leaves us
a large freedom to suit phenomenological needs.\par
We have also examined the possibility of using kernel-based 3D reductions
in the two and three-fermion problems. We found that this kind of 3D
reduction can also be used in the calculation of bound states, despite the
fact that they do not lead to the correct scattering operators (excepting
in Phillips and Wallace's reduction).\par
As a first test of our methods in a specific problem, we are presently
working on a system of two or three fermions exchanging photons in Feynman or
Coulomb's gauges \cite{38*}. The various terms generated by our
propagator-based 3D reduction will be compared with these generated by a
kernel-based 3D reduction and with these written in the framework of
Gross' spectator model.

\end{document}